\pdfoutput=1
\documentclass[prl,twocolumn,superscriptaddress,longbibliography]{revtex4-2}
\usepackage{graphicx,amsmath,amssymb,tikz,scalerel,hyperref}
\hypersetup{
        colorlinks=true,       
        linkcolor=red,        
        citecolor=blue,     
        urlcolor=blue}
\renewcommand{\section}[1]{{\par\it #1.---}\ignorespaces}
\usetikzlibrary{svg.path}
\definecolor{orcidlogocol}{HTML}{A6CE39}
\tikzset{
	orcidlogo/.pic={
		\fill[orcidlogocol] svg{M256,128c0,70.7-57.3,128-128,128C57.3,256,0,198.7,0,128C0,57.3,57.3,0,128,0C198.7,0,256,57.3,256,128z};
		\fill[white] svg{M86.3,186.2H70.9V79.1h15.4v48.4V186.2z}
		svg{M108.9,79.1h41.6c39.6,0,57,28.3,57,53.6c0,27.5-21.5,53.6-56.8,53.6h-41.8V79.1z M124.3,172.4h24.5c34.9,0,42.9-26.5,42.9-39.7c0-21.5-13.7-39.7-43.7-39.7h-23.7V172.4z}
		svg{M88.7,56.8c0,5.5-4.5,10.1-10.1,10.1c-5.6,0-10.1-4.6-10.1-10.1c0-5.6,4.5-10.1,10.1-10.1C84.2,46.7,88.7,51.3,88.7,56.8z};}}
\newcommand\orcid[1]{\href{https://orcid.org/#1}{\mbox{\scalerel*{\begin{tikzpicture}[yscale=-1,transform shape]\pic{orcidlogo};\end{tikzpicture}}{|}}}}

\begin{document}
\title{Kerr nonlinearity induced strong spin-magnon coupling}
\author{Feng-Zhou Ji\orcid{0000-0003-4859-1535}}
\affiliation{Key Laboratory of Quantum Theory and Applications of MoE, Lanzhou Center for Theoretical Physics, and Key Laboratory of Theoretical Physics of Gansu Province, Lanzhou University, Lanzhou 730000, China}
\author{Jun-Hong An\orcid{0000-0002-3475-0729}}
\email{anjhong@lzu.edu.cn}
\affiliation{Key Laboratory of Quantum Theory and Applications of MoE, Lanzhou Center for Theoretical Physics, and Key Laboratory of Theoretical Physics of Gansu Province, Lanzhou University, Lanzhou 730000, China}

\begin{abstract}
One pillar of quantum magnonics is the exploration of the utilization of the mediation role of magnons in different platforms to develop quantum technologies. The efficient coupling between magnons and various quantum entities is a prerequisite.  Here, we propose a scheme to enhance the spin-magnon coupling by the magnonic Kerr nonlinearity in a YIG sphere. We find that the Kerr-enhanced spin-magnon coupling invalidates the widely used single-Kittel-mode approximation to magnons. It is revealed that the spin decoherence induced by the multimode magnons in the strong-coupling regime becomes not severe, but suppressed, manifesting as either population trapping or persistent Rabi-like oscillation. This anomalous effect is because the spin changes to be so hybridized with the magnons that one or two bound states are formed between them. Enriching the spin-magnon coupling physics, the result supplies a guideline to control the spin-magnon interface.
\end{abstract}
\maketitle

\section{Introduction}\label{sec:level1}
Magnons are the elementary excitation of a collective spin wave in magnetic materials. The quantized interactions of magnons with different quantum platforms have inspired many novel applications in quantum technologies \cite{YUAN20221,Lachance-Quirion_2019,ZARERAMESHTI20221,doi:10.1126/sciadv.1501286,10.1038/ncomms5700,PhysRevLett.125.027201,PhysRevLett.127.087203,PhysRevLett.124.053602,PhysRevLett.102.177206,10.1038/s41467-022-35174-9,10.1063/5.0121314}. Besides quantum transduction \cite{PhysRevB.93.174427}, memory \cite{Zhang2015}, sensing \cite{PhysRevApplied.13.064001,Crescini2020}, and unidirectional invisibility \cite{PhysRevLett.123.127202}, using the coupling between magnons and photons or phonons, the efficient couplings of magnons to spins have attracted much attention due to their potential realization of quantum networks \cite{PhysRevA.100.022343} and quantum sensing \cite{doi:10.1126/science.aaz9236,PhysRevResearch.4.L012025,PhysRevApplied.16.064058}. The efficient spin-magnon couplings via either direct interactions \cite{doi:10.1126/science.aaa3693,PhysRevLett.125.247702,PRXQuantum.2.040314,PhysRevB.105.075410,PhysRevResearch.4.023221,PhysRevResearch.4.043180,PhysRevX.12.011060,PhysRevLett.129.037205} or the indirect way by exchanging photons or phonons \cite{PhysRevA.103.052411,PhysRevA.103.063702,PhysRevA.107.033516,PhysRevLett.130.073602,PhysRevA.107.033516} have been proposed. How to enhance the spin-magnon coupling strength is a prerequisite to explore their applications.

The Kerr nonlinearity of magnons in magnetic materials supplies a useful mechanism in quantum-state engineering \cite{Moslehi_2023,10.1007/s11433-018-9344-8}. Based on it, magnon-polariton bistability \cite{PhysRevLett.120.057202,PhysRevResearch.3.023126,PhysRevLett.129.123601} and tristability \cite{PhysRevLett.127.183202,PhysRevB.103.104411}, which are useful in the microwave nonreciprocal transmission \cite{PhysRevApplied.12.034001}, high-order sideband \cite{10.1364/OL.43.003698,PhysRevA.104.033708,PhysRevApplied.18.044074} and entanglement generations \cite{PhysRevResearch.1.023021}, and quantum phase transition \cite{PhysRevB.106.054419}, have been reported. A scheme to enhance the spin-magnon coupling using the Kerr nonlinearity was proposed in Ref. \cite{PhysRevB.105.245310}. These works on quantum magnonics were generally based on an approximation in which the magnons are effectively treated as a single first-order Kittel mode \cite{PhysRevLett.121.203601,PhysRevB.101.060407,PhysRevA.104.032606,PhysRevB.105.094422}. It has been revealed that the higher-order magnonic modes are non-negligible in the presence of the Kerr nonlinearity \cite{10.1038/s41598-017-11835-4,PhysRevLett.130.046705,PhysRevB.107.144411}. References \cite{PRXQuantum.2.040314,PhysRevResearch.4.023221,PhysRevB.105.075410} studied the interactions between spins and multimode magnons. However, they are under the Markovian approximation, which is only valid in the weak-coupling condition.

\begin{figure}[tbp]
\includegraphics[width=.9\columnwidth]{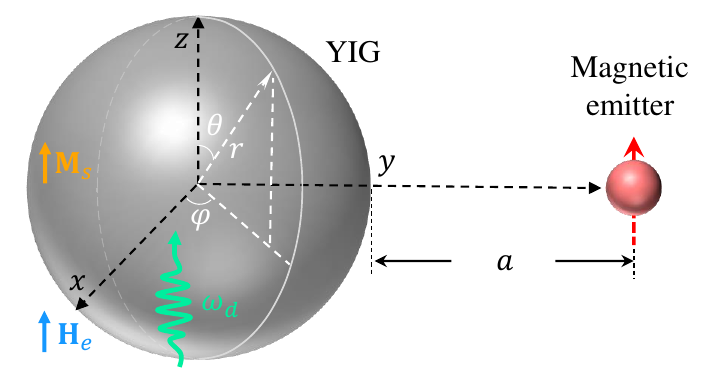}
\caption{ Schematic illustration of the system. A magnetic emitter interacts with the magnons in a YIG sphere with radius $R$ in a static magnetic field $\textbf{H}_e$, and an induced magnetic field $\textbf{M}_s$. A driving field with frequency $\omega_d$ is applied. }\label{sketch}
\end{figure}

Here, we investigate the non-Markovian dynamics of a spin defect coupled to magnons in a YIG sphere. A scheme to enhance the spin-magnon coupling by the magnonic Kerr nonlinearity is proposed. We find that the increasing of the coupling invalidates the widely used single-mode approximation of magnons to describe the matter-magnon coupling. The strong coupling also causes the spin to exhibit features with a suppressed decoherence, i.e., from the conventional oscillating damping to either the population trapping or the persistent Rabi-like oscillation. Our analysis reveals that such an anomalous decoherence is due to the formation of different numbers of spin-magnon bound states. Indicating that the Kerr nonlinearity endows the spin-magnon interface with a good controllability, our result paves the way to design quantum magnon devices.

\section{System and spectral density}
We consider a spin defect as a magnetic emitter coupled to the magnons attached to a YIG sphere in the presence of the Kerr nonlinearity (see Fig. \ref{sketch}). Its Hamiltonian reads \cite{PhysRevLett.125.247702,PhysRevResearch.4.043180}
\begin{eqnarray}
\begin{aligned}
\hat{\mathcal H}_\text{Kerr}=&\hbar\omega_{0}\hat{\sigma}^\dag\hat{\sigma}+\sum_{k}\big[\hbar\omega_{k}\hat{b}_{k}^{\dagger}\hat{b}_{k}-(\hbar K/2)\hat{b}_{k}^{\dagger2}\hat{b}_{k}^2\\
&-(g_k\hat{b}_{k}^{\dagger}\hat{\sigma}-\Omega_{d}e^{i\phi_d}\hat{b}_{k}^{\dagger}e^{-i\omega_{d}t}+\text{H.c.})\big].\label{H1}
\end{aligned}
\end{eqnarray}
Here, $\hat{\sigma}=|g\rangle\langle e|$ is the transition operator of the spin defect with frequency $\omega_{0}$ from the excited state $|e\rangle$ to the ground state $|g\rangle$, $\hat{b}_{k}$ is the annihilation operator of the $k$th magnon mode with frequency $\omega_{k}$, and $g_k=\mu_{0}\textbf{m}\cdot\tilde{\textbf{H}}_{k}^{\ast}(\textbf{r})$, with $\textbf{m}$ being the spin magnetic moment and $\tilde{\textbf{H}}_{k}(\textbf{r})$ being the vacuum amplitude of the $k$th magnon mode concentrated around the YIG sphere, is their coupling strength. An inhomogeneous rf magnetic excited field is needed to trigger the multimode magnons \cite{10.1063/1.1723117}. The magnons are further driven by a microwave field with frequency $\omega_{d}$, amplitude $\Omega_{d}$, and phase  $\phi_d$. The Kerr nonlinearity quantified by $K$ is caused by the magnetocrystalline anisotropy of the YIG sphere. It has been used to generate the magnon squeezing \cite{10.1007/s11433-018-9344-8}, which is attractive to the application of the magnons \cite{YUAN20221,PhysRevB.101.054402}. Rewriting the magnon operators as the sum of their steady-state mean value and fluctuations, i.e., $\hat{b}_{k}=\langle \hat{b}_{k}\rangle+\delta\hat{b}_{k}$, and neglecting the high-order fluctuation terms in the strong driving condition, Eq. \eqref{H1} in the rotating frame with $\hat{\mathcal H}_0=\omega_d(\hat{\sigma}^\dag\hat{\sigma}+\sum_{k}\hat{b}_{k}^{\dagger}\hat{b}_{k})$ becomes (see Supplemental Material \cite{SM})
\begin{equation}
\begin{aligned}
\hat{\mathcal H}_\text{Linear}=&\hbar\Delta_0\hat{\sigma}^\dag\hat{\sigma}+\sum_{k}\big\{\hbar(\omega_{k}-\Pi_k)\hat{b}_{k}^{\dagger}\hat{b}_{k}\\
&-[g_{k}\hat{b}_{k}^{\dagger}\hat{\sigma}+(\hbar\mathcal{K}_k/2)\hat{b}_{k}^{2}+\text{H.c.}]\big\},\end{aligned}\label{had}
\end{equation}
where $\Delta_0=\omega_{0}-\omega_{d}$, $\Pi_k=\omega_{d}+2\mathcal{K}_k$, and $\mathcal{K}_k= K\langle \hat{b}_{k}\rangle^{2}$. We have rewritten $\delta\hat{b}_{k}$ as $\hat{b}_{k}$ for brevity. Making the Bogoliubov transformation $\hat{\mathcal S}=\exp[\sum_kr_k(\hat{b}_k^2-\hat{b}_k^{\dag2})/2]$, with $r_{k}=\frac{1}{4}\ln(\frac{\omega_{k}-\Pi_k+\mathcal{K}_k}{\omega_{k}-\Pi_k-\mathcal{K}_k})$, to Eq. \eqref{had} and neglecting the counter-rotating terms \cite{PhysRevB.105.245310}, we obtain
\begin{equation}
\hat{\mathcal H}=\hbar\Delta_0\hat{\sigma}^\dag\hat{\sigma}+\sum_{k}[\hbar\zeta_{k}\hat{b}_{k}^{\dagger}\hat{b}_{k}-(\mathcal{G}_{k}
\hat{b}_{k}^{\dagger}\hat{\sigma}+\text{H.c.})],\label{H3}
\end{equation}
where $\zeta_{k}=(\omega_{k}-\Pi_k)/\cosh{(2r_{k})}$ and $\mathcal{G}_{k}=e^{r_{k}}g_{k}/2$. We find that the spin-magnon coupling is exponentially enhanced and the magnon frequencies are suppressed by the Kerr nonlinearity assisted by the microwave driving. This is one of our main results. To simplify our discussion, we approximate $\Pi_k\simeq\omega_d$ due to $\omega_d\gg \mathcal{K}_k$ and $r_k\equiv r$ by neglecting their $k$-dependence, which is valid by properly choosing $\Omega_d$ and $\omega_d$ in our finite magnonic bandwidth case \cite{10.1063/1.1723117}.

Consider that the YIG sphere is at low temperature such that the magnons are initially in the vacuum state \cite{PhysRevResearch.1.023021,PhysRevB.104.064426}. After tracing over the magnonic degrees of freedom from the dynamics of the spin-magnon system, we derive an exact master equation of the spin as (see Supplemental Material \cite{SM})
\begin{equation}
\dot{\rho}(t)=i\Omega(t)[\rho(t),\hat{\sigma}^\dag\hat{\sigma}]+\Gamma(t)[2\hat{\sigma}\rho(t)\hat{\sigma}^\dag
-\{\hat{\sigma}^\dag\hat{\sigma},\rho(t)\}].\label{exactME}
\end{equation}
The renormalized frequency is $\Omega(t)=-\text{Im}[\dot{c}(t)/c(t)]$ and the decay rate is $\Gamma(t)=-\text{Re}[\dot{c}(t)/c(t)]$, where $c(t)$ satisfies
\begin{equation}
\dot{c}(t)+i\Delta_{0}c(t)+\int_{0}^{t}d\tau c(\tau)f(t-\tau)=0,\label{dynamics111}
\end{equation}
under $c(0)=1$. The convolution in Eq. \eqref{dynamics111} makes the dynamics non-Markovian with all the memory effects incorporated in the time-dependent coefficients in Eq. \eqref{exactME}. The magnonic correlation function is $f(t-\tau)=\int_{\zeta_{\text{min}}}^{\zeta_{\text{max}}} d\zeta J(\zeta)e^{-i\zeta(t-\tau)}$ and the spectral density is \cite{PhysRevLett.125.247702}
\begin{equation}
J(\zeta)=\frac{\eta\mu_{0}}{4\hbar\pi}\text{Im}[\textbf{m}^{\ast}\cdot k^2\textbf{G}(\textbf{r},\textbf{r},\zeta \cosh(2r)+\Pi)\cdot\textbf{m}].\label{spectrumdensity11}
\end{equation}
where $k=[\zeta\cosh(2r)+\Pi]/c$ and $\eta=e^{2r}\cosh{(2r)}$. The Green's tensor reads
$\bar{k}^{2}\textbf{G}(\textbf{r}, \textbf{a},\bar{\omega})=\sum_{\alpha,\beta\in \{r,\theta,\varphi\}}[H_{0,\alpha}^{\beta}+H_{\alpha}^{\beta}]\textbf{e}_{\alpha}\textbf{e}_{\beta}$, where ${\bf H}=-{\pmb \nabla}\phi$ and $\bf{H}_0$ takes the similar form as ${\bf H}$ but in the absence of the YIG sphere. The potential $\phi$ caused by the YIG satisfying $\phi={1\over 4\pi}{\pmb \nabla}_{ a}{1\over |{\bf r}-{\bf a}|}$ is subject to the boundary condition of $(1+\chi)(\frac{\partial^{2}}{\partial x^{2}}+\frac{\partial^{2}}{\partial y^{2}})\phi+\frac{\partial^{2}\phi}{\partial z^{2}}=0$ in $r\leq R$ and $\nabla^2\phi=0$ in $r>R$ \cite{Mills1974}. $\pmb{\chi }$ is the magnetic susceptibility tensor and determined by the Landau-Lifshitz-Gilbert equation as \cite{PhysRev.105.390,10.1063/1.1735216}
\begin{equation}
\begin{aligned}
\chi_{xx}&=\chi_{yy}=\frac{\gamma^{2}h_{0}M_{s}}{\gamma^{2}h_{0}^{2}-\omega^{2}-i\Gamma_{0}\omega}\equiv\chi,\\
\chi_{xy}&=\chi^{\ast}_{yx}=i\frac{\gamma\omega M_{s}}{\gamma^{2}h_{0}^{2}-\omega^{2}-i\Gamma_{0}\omega}\equiv i\kappa,\label{magsus}
\end{aligned}
\end{equation}
where $\gamma$ is the gyromagnetic ratio, $\Gamma_{0}= 2\gamma h_{0}\alpha$, with $\alpha$ being the Gilbert parameter, is the damping parameter, $\textbf{h}_{0}=h_{0}\textbf{e}_{z}=\textbf{H}_{e}+\textbf{H}_{d}$, with $\textbf{H}_{e}$ being the external static field and $\textbf{H}_{d}=-\textbf{M}_{s}/3$ being the demagnetization field, and ${\bf M}_s$ is the saturation magnetization. Putting the spin on the equatorial plane of the YIG sphere, i.e., $\theta=\pi/2$, and choosing $\textbf{m}=-\mu_{B}({\bf e}_{x}+i{\bf e}_{y})=-\mu_B e^{i\varphi}({\bf e}_r+i{\bf e}_\varphi)$, we have the nonzero components of the Green's tensor as $\text{Im}[\textbf{m}^{\ast}\cdot\textbf{G}\cdot\textbf{m}]=\mu_B^2[\text{Im}(\textbf{G}_{rr}+\textbf{G}_{\varphi\varphi})+\text{Re}(\textbf{G}_{r\varphi}-\textbf{G}_{\varphi r})]$ \cite{PhysRevB.101.060407,PhysRevLett.125.247702,PhysRevB.105.075410}.
The analytic form of $\textbf{G}(\textbf{r},\textbf{a},\bar{\omega})$ is given in (see Supplemental Material \cite{SM}). The Green's tensor and the spectral density show resonance peaks determined by \cite{PhysRev.105.390,10.1063/1.1735216}
\begin{equation}
(n+1-m\kappa)P_{n}^{m}(\xi_0)+\xi_{0}P_{n}^{m{\prime}}(\xi_{0})=0,\label{seculareq}
\end{equation}
where $P_{n}^{m}$ is the associated Legendre polynomial and $\xi_{0}=(1+1/\chi)^{1/2}$. The magnon modes corresponding to $n=-m=1$, $2$, and $3$ in the absence of the Kerr nonlinearity are the dipole or Kittel mode $\omega_{\text{K}}=\gamma(h_{0}+M_s/3)$, the quadrupolar mode $\omega_{\text{Q}}=\gamma(h_{0}+2M_s/5)$, and the octupolar mode $\omega_{\text{O}}=\gamma(h_{0}+3M_s/7)$, respectively. In the presence of the Kerr nonlinearity, they become $\zeta_{\text{K,Q,O}}=(\omega_{\text{K,Q,O}}-\Pi)/\cosh(2r)$. Ranging all $m$ and $n$, it was found that the frequency range of $J(\zeta)$ is from $\zeta_{\text{min}}=(\gamma h_{0}-\Pi)/\cosh(2r)$ to $\zeta_{\text{max}}=[\gamma (h_{0}+M_{s}/2)-\Pi]/\cosh(2r)$ \cite{10.1063/1.1723117}.

\section{Spin dynamics}
A widely used approximation in studying matter-magnon coupling is the Markovian approximation \cite{PhysRevB.104.134423,PRXQuantum.2.040314,PhysRevLett.121.203601,PhysRevB.105.094422,PhysRevB.105.075410,PhysRevB.103.224401,PhysRevB.104.064423,PhysRevResearch.4.023221}. It is valid when their coupling is weak and the time scale of $f(t-\tau)$ is much smaller than the one of the matter. After replacing $c(\tau)$ by $c(t)$ and extending the upper bound of the time integral to infinity, the Markovian approximate solution of Eq. \eqref{dynamics111} is $c_{\text{MA}}(t)=e^{-[\Lambda+i\Upsilon(\Delta_0)]t}$, where $\Upsilon(\Delta_0)=\mathcal{P}\int d\zeta J(\zeta)/(\Delta_{0}-\zeta)$ is the Lamb shift and $\Lambda=\pi J(\Delta_{0})$ is the spontaneous emission rate. The exponential-decay feature of $|c_\text{MA}(t)|^2$ characterizes a unidirectional energy flow from the spin to the magnons and the destructive effect of the magnons on the spin. This approximation cannot reflect the energy back flow induced by the strong spin-magnon coupling  \cite{PhysRevLett.125.247702,PhysRevResearch.4.043180}.

A pseudocavity method was proposed to study the strong light-matter coupling in an absorptive medium \cite{PhysRevLett.112.253601,PhysRevB.89.041402,PhysRevB.95.161408,PhysRevLett.120.030402,PhysRevLett.126.093601}. Keeping only the Kittel mode $\zeta_\text{K}$, we approximate $J(\zeta)$ as a Lorentzian form $\mathcal{J}(\omega)=\frac{J(\zeta_\text{K})(\gamma_{p}/2)^2}{(\omega-\zeta_\text{K})^{2}+(\gamma_{p}/2)^{2}}$. The system is effectively seen as a spin coherently interacting with a pseudocavity mode $\hat{a}$ with frequency $\zeta_\text{K}$ and damping rate $\gamma_p$ in a coupling strength $g^2=\pi J(\zeta_\text{K})\gamma_{p}/2$. Here, $\gamma_{p}$ is relevant to the damping parameter $\Gamma_0$. Thus, the spin dynamics is phenomenologically described by
\begin{equation}
\begin{aligned}
\dot{\rho}(t)=&i[\rho(t),\Delta_{0}\hat{\sigma}^\dag\hat{\sigma}+\zeta_\text{K}\hat{a}^{\dagger}\hat{a}+g(\hat{a}\hat{\sigma}^\dag+\text{H.c.})]\\
&+\frac{\gamma_{p}}{2}[2\hat{a}\rho(t)\hat{a}^{\dagger}-\{\hat{a}^{\dagger}\hat{a},\rho(t)\}].\label{pseudocavity}
\end{aligned}
\end{equation}
Although partially reflecting the energy back flow from the magnons to the spin, this method misses important physics from the magnonic higher-order resonant modes.

To fully capture the physics of the strong spin-magnon coupling enhanced by the Kerr nonlinearity and uncover the condition under which the pseudocavity method is applicable, we investigate the exact spin dynamics by choosing $\Delta_0=\zeta_\text{K}$. The steady-state solution of  Eq. \eqref{dynamics111} is computable by a Laplace transform. It converts Eq. \eqref{dynamics111} into $\tilde{c}(s)=[s+i\Delta_{0}+\int_{\zeta_\text{min}}^{\zeta_\text{max}} d\zeta\frac{J(\zeta)}{s+i\zeta}]^{-1}$. $c(t)$ is obtained by making an inverse Laplace transform to $\tilde{c}(s)$, which requires finding its poles via (see Supplemental Material \cite{SM})
\begin{equation}
{E\over\hbar}=\Delta_{0}+\int_{\zeta_\text{min}}^{\zeta_\text{max}}\frac{J(\zeta)}{E/\hbar-\zeta}d\zeta\equiv Y(E),\label{cs11}
\end{equation}
where $E=i\hbar s$. First, the roots $E$ of Eq. \eqref{cs11} are exactly the eigenenergies of the total spin-magnon system.
To prove this, we expand the eigenstate as $|\phi_{E}\rangle=x|e,\{0_k\}\rangle+\sum_{k}y_{k}|g,1_{k}\rangle$. Substituting $|\phi_{E}\rangle$ into $\hat{\mathcal H}|\phi_{E}\rangle=E|\phi_{E}\rangle$, we readily obtain Eq. \eqref{cs11}. Second, because $Y(E)$ is a decreasing function in the regimes $E\in (-\infty,\hbar\zeta_\text{min}]$ and $[\hbar\zeta_\text{max},+\infty)$, Eq. \eqref{cs11} has one isolated root $E^\text{b}$ in $(-\infty,\hbar\zeta_\text{min}]$ or $[\zeta_\text{max},+\infty)$ provided $Y(\hbar\zeta_\text{min})<\hbar\zeta_\text{min}$ or $Y(\hbar\zeta_\text{max})>\hbar\zeta_\text{max}$. The eigenstate corresponding to $E^\text{b}$ is called the bound state. On the other hand, $Y(E)$ is non-analytical in the regime $E\in[\hbar\zeta_\text{min},\hbar\zeta_\text{max}]$ due to the singularity in its integration. Therefore, Eq. \eqref{cs11} has an infinite number of roots in this regime, which form an energy band. Using the residue theorem, we have $c(t)=\sum_{j=1}^MZ_je^{{-i\over\hbar}E^b_jt}+\int_{\zeta_\text{min}}^{\zeta_\text{max}}\Theta(E)e^{-iEt}dE$, where $\Theta(E)={J(E)\over[E-\Delta_0-\Upsilon(E/\hbar)]^2+[\pi J(E)]^2}$, $M$ being the number of the bound states, and $Z_{j}=[1+\int_{\zeta_\text{min}}^{\zeta_\text{max}}\frac{J(\zeta)d\zeta}{(E^\text{b}_{j}/\hbar-\zeta)^2}]^{-1}$ the residue contributed by the $j$th bound state. Oscillating with time in continuously changing frequencies ${E/\hbar}$ of the band energies, the integrand tends to zero in the long-time limit due to the out-of-phase interference. Thus, the steady-state solution of Eq. \eqref{cs11} is \cite{PhysRevB.106.115427}
\begin{equation}
\lim_{t\rightarrow\infty}c(t)=\begin{cases}0, ~~~~~~~~~~~~~~~~~~~~\text{no bound state}\\ \sum_{j=1}^MZ_je^{{-i\over\hbar}E^\text{b}_jt}, ~M~\text{bound states}\end{cases}\hspace{-.3cm}.\label{longtimesolution111}
\end{equation}
Assuming the spin is initially in $|e\rangle$ and solving Eq. \eqref{exactME}, we obtained that the excited-state population is just $|c(t)|^2$ (see Supplemental Material \cite{SM}). Thus, Eq. \eqref{longtimesolution111} reveals that thanks to the Kerr-nonlinearity-enhanced spin-magnon coupling, the formation of the bound states would prevent the spin from relaxing to its ground state. Since it is not obtained from both the Markovian approximation and the pseudocavity method, such an anomalous decoherence manifests the distinguished role played by the non-Markovian effect and the feature of the energy spectrum of the total spin-magnon system in the decoherence of the spin. This is another main result of our work.

\begin{figure}[tbp]
\includegraphics[width=1.0\columnwidth]{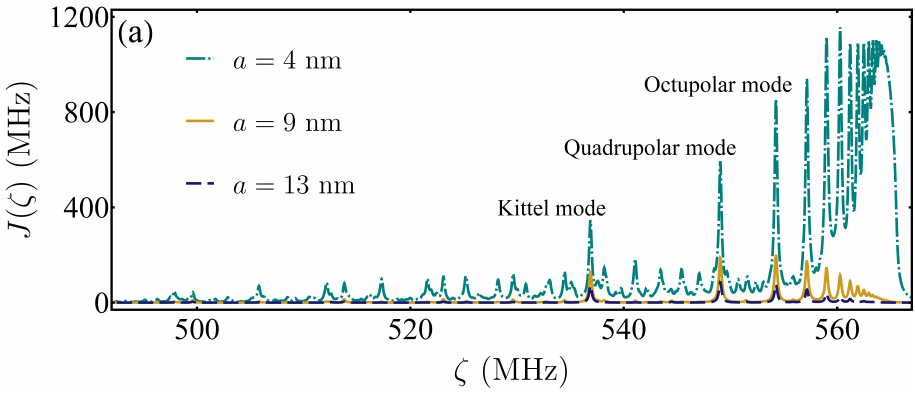}\\
\includegraphics[width=\columnwidth]{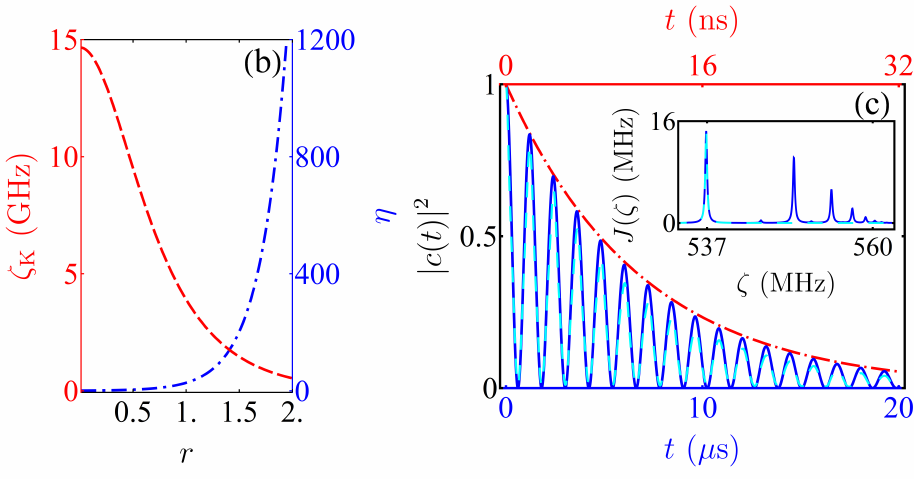}\\
\caption{(a) Spectral density $J(\zeta)$ in different spin-YIG distance $a$. (b) Kittel-mode frequency $\zeta_\text{K}$ and enhancement coefficient $\eta$ in different $r$. (c) Evolution of the excited-state population $|c(t)|^2$ from the Markovian approximation (red dotted line), the pseudocavity method (cyan dashed line), and the non-Markovian dynamics (blue line). The inset is $J(\zeta)$ and the fitted Lorentzian form $\mathcal{J}(\zeta)$. We use $\omega_{d}= 1$ GHz, $\gamma= 28~\text{GHz} \cdot \text{T}^{-1}$, $\Gamma_{0}=8\times 10^{-3}$ GHz, $M_{s}=0.178$ T, $h_0=0.5$ T, $R=30$ nm, $r=2$ in (a) and (c), and $a=26 $ nm in (c).} \label{spectrum}
\end{figure}

\section{Numerical results}\label{sec:withdissipation}
We plot in Fig. \ref{spectrum}(a) the spectral density $J(\zeta)$ in different spin-YIG distance $a$ for $r=2$. The driving-field frequency $\omega_{d}$ is chosen as $\omega_{d}= 1$ GHz and the Kerr coefficient $K$ relates to the volume $V$ of the three-dimensional YIG sphere, i.e., $K\varpropto V^{-1}$ \cite{10.1007/s11433-018-9344-8} and is about kHz \cite{PhysRevB.105.245310}, which makes the validity of $\Pi\simeq \omega_d$. We really see that $J(\zeta)$ exhibits obvious peaks at $\zeta=537$, $549$, and $554$ MHz irrespective of the value of $a$, which match well with our analytical frequencies $\zeta_{\text{K,Q,O}}$ of the Kittel, quadrupolar, and octupolar modes evaluated from Eq. \eqref{seculareq}. With decreasing $a$, $J(\zeta)$ shows an increase due to the near-field enhancement \cite{PhysRevResearch.4.043180}. It signifies a strong spin-magnon coupling in the small-$a$ regime.  Figure \ref{spectrum}(b) shows the effect of the Kerr nonlinearity on enhancing the spin-magnon coupling. It reveals that, with increasing $r$ from zero to 2, $\zeta_{\text{K}}$ decreases from $15$ GHz to $537$ MHz, while the prefactor $\eta$ of $J(\zeta)$ increases from $1$ to $1200$. An efficient increase of about four orders of magnitude of $\eta/\zeta_\text{K}$ manifests a dramatic boost of the spin-magnon coupling strength. It confirms that the Kerr nonlinearity can be used to enhance the spin-magnon coupling \cite{PhysRevB.105.245310,PhysRevLett.129.123601,PhysRevLett.127.183202}.

\begin{figure}[tbp]
\includegraphics[width=\columnwidth]{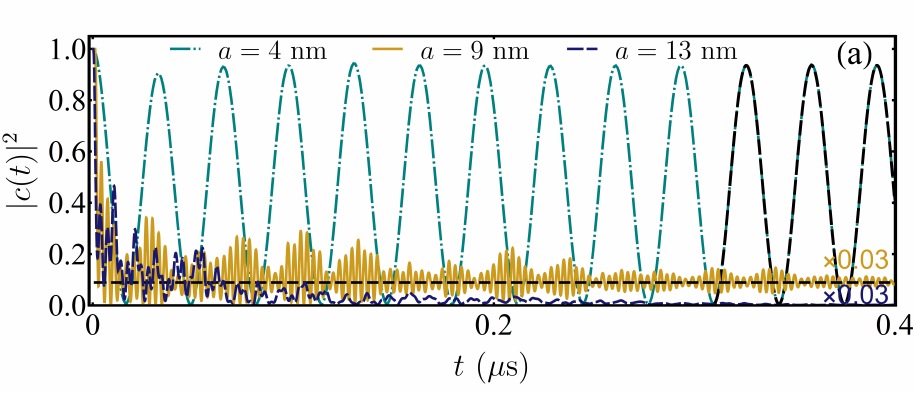}\\
\includegraphics[width=\columnwidth]{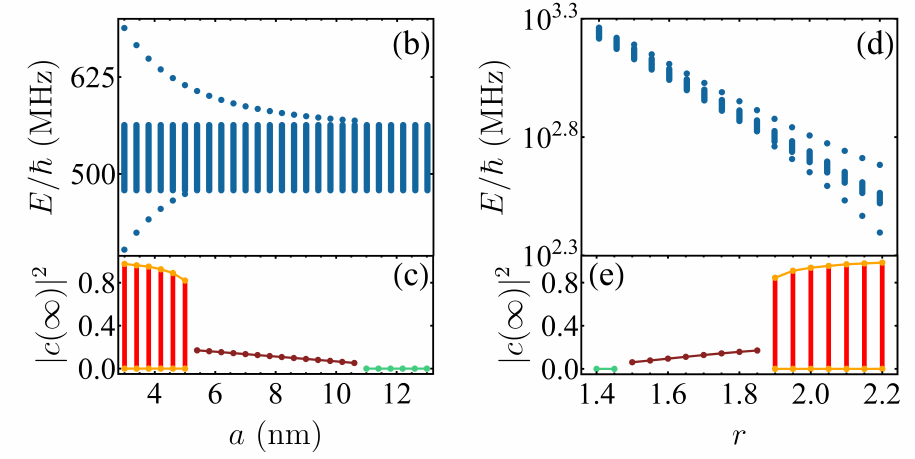}\\
\includegraphics[width=\columnwidth]{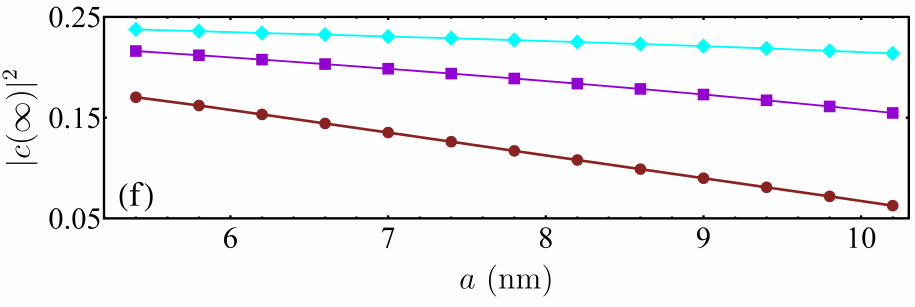}
\caption{ (a) Evolution of $|c(t)|^2$ in different $a$. The black dashed lines are the corresponding steady-state values from Eq. \eqref{longtimesolution111}. The time for the cases of $a=9$ and $13$ nm is magnified by a factor of 0.03. Energy spectrum of the whole system in different (b) $a$ and (d) $r$ obtained by solving Eq. \eqref{cs11}. $|c(\infty)|^2$ from solving Eq. \eqref{dynamics111} denoted by the dots and from Eq. \eqref{longtimesolution111} denoted by the solid lines in different (c) $a$ and (e) $r$. The red region covers the values during its persistent oscillation. (f) $|c(\infty)|^2$ when $\Delta_0=\zeta_\text{K}$ (brown dot), $\zeta_\text{Q}$ (purple square), and $\zeta_\text{O}$ (cyan rhombus). $r=2.0$ in (b) and (c), $a=4$ nm in (d) and (e), and others are the same as Fig. \ref{spectrum}(c).}\label{dyn}
\end{figure}

Figure \ref{spectrum}(c) shows the comparison of $|c(t)|^2$ obtained by three methods when $a=26$ nm. The exponential decay in the Markovian result entirely fails to describe the rapid spin-magnon energy exchanges obtained via numerically solving Eq. \eqref{dynamics111}, which is fully captured by the pseudocavity method. It is the signature of the non-Markovian memory effect owned by the strong-coupling dynamics \cite{acsphotonics2018}. In this case, $J(\zeta)$ is dominated by the Kittel mode such that a Lorentzian fitting centered at $\zeta_\text{K}$ is sufficient and the pseudocavity method works well. However, with further decreasing $a$ [see Fig. \ref{spectrum}(a)] or increasing $r$, the high-order magnon modes become dominated, where the pseudocavity method no longer work.

Figure \ref{dyn}(a) shows the exact $|c(t)|^2$ in different $a$ when $r=2$. The strong spin-magnon coupling favored by both the near-field enhancement \cite{PhysRevResearch.4.043180} and Kerr nonlinearity causes $|c(t)|^2$ to exhibit rich behaviors. It is interesting to find that, in contrast to the damping to zero for $a=13$ nm, which has no qualitative difference from the result predicted by the pseudocavity method, $|c(t)|^2$ approaches a finite value when $a=9$ nm, while it exhibits a lossless Rabi-like oscillation when $a=4$ nm. It reveals an anomalous behavior in which a small spin-magnon distance with the Kerr nonlinearity induces a strong spin-magnon coupling, which, on the contrary, causes a suppressed decoherence. It is not expected that a stronger spin-magnon coupling always causes a more severe decoherence to the spin \cite{PhysRevLett.130.073602,PhysRevB.100.235453}. This behavior can be explained by the features of the energy spectrum of the total spin-magnon system. Figure \ref{dyn}(b) indicates that two branches of bound states separate the energy spectrum into three regimes. When $a\geq10.4$ nm, no bound state is formed and thus $|c(t)|^2$ decays to zero. When $4.8$ nm $<a<$ $10.4$ nm, one bound state is present and $|c(t)|^2$ tends to finite values. When $a\leq 4.8$ nm, two bound states are present and $|c(t)|^2$ behaves as a persistent Rabi-like oscillation in a frequency $|E^\text{b}_1-E^\text{b}_2|/\hbar$. The matching of the long-time behaviors of the three regimes with the analytical result in Eq. \eqref{longtimesolution111} verifies the distinguished role played by the bound states and non-Markovian effect in determining the strong-coupled spin-magnon physics; see Fig. \ref{dyn}(c). Figures \ref{dyn}(d) and \ref{dyn}(e) indicate that it is just the Kerr-nonlinearity-induced strong spin-magnon coupling that causes the formation of the bound states and the accompanying population trapping and persistent Rabi-like oscillation. Figure \ref{dyn}(f) shows $|c(\infty)|^2$ in the distances $a$ supporting the formation of one bound state. It demonstrates that the trapped population $|c(\infty)|^2$ can be controlled by choosing $\Delta_0$ as different magnonic resonant frequencies. All the results prove that the strong spin-magnon coupling endows the spin with rich anomalous decoherence governed by the formation of different numbers of bound states. It supplies a guideline to control the spin coherence via engineering the feature of the spin-magnon energy spectrum.

\section{Discussion and conclusions}\label{sec:Conclusion}
Quantum magnonics exploring the efficient couplings between magnons and different kinds of quantum matter has made great progress \cite{doi:10.1126/sciadv.1501286,PhysRevLett.113.083603,RN1,PhysRevLett.111.127003,PhysRevB.94.224410,PhysRevLett.120.057202,PhysRevLett.128.047701,PhysRevResearch.4.L012025,YUAN20221,PhysRevLett.129.123601,PhysRevLett.130.073602,PhysRevA.107.033516}. Many of these works were based on the single-magnon-mode approximation, which may be insufficient in the strong matter-magnon coupling.
The coupling between single spins and multimode magnons was studied in Ref. \cite{PhysRevLett.125.247702}, but the Kerr nonlinearity was absent. The Kerr nonlinearity has been observed in cavity magnon mechanics formed by the YIG \cite{PhysRevLett.120.057202,PhysRevB.94.224410,PhysRevLett.129.123601}.  The bound state and its distinguished role in the non-Markovian dynamics have been experimentally observed in both photonic crystal \cite{Liu2017} and ultracold-atom \cite{Krinner2018,Kwon2022} systems. The progresses give a strong support that our finding is realizable in state-of-the-art experiments \cite{PhysRevApplied.19.014002,PhysRevLett.128.047701}. Note that, although only the YIG is studied, our results are applicable to other magnetic materials, such as CoFeB \cite{PhysRevB.104.214416,PhysRevResearch.4.L012025}. As a final remark, the expectation value of the magnonic fluctuation operator described by $\hat{b}_k$ in Eq. \eqref{H3} should be zero to ensure the self-consistence of our linearization approximation to Eq. \eqref{H1}. This can be proven as follows. The evolved state of the spin-magnon system under our studied initial condition $|\Psi_\text{tot}(0)\rangle=|e,\{0_k\}\rangle$ reads $|\Psi_\text{tot}(t)\rangle=c(t)|e,\{0_k\}\rangle+\sum_kd_k(t)|g,1_k\rangle$, which readily leads to $\langle\Psi_\text{tot}(t)|\hat{b}_k|\Psi_\text{tot}(t)\rangle=0$.

In summary, we have investigated the near-field interactions between a spin defect and magnons in a YIG sphere with Kerr nonlinearity. It is found that the Kerr nonlinearity induces a dramatic enhancement to the spin-magnon coupling. Contrary to the belief that a strong coupling always causes a severe decoherence, such a strong coupling makes the magnon-induced decoherence to the spin change from complete damping to either population trapping or persistent Rabi-like oscillation. This anomalous decoherence is due to the formation of different numbers of spin-magnon bound states. Breaking the dissipation barrier of the spin, our finding supplies a guideline to suppress the spin decoherence and design quantum magnonic devices.

\section{Acknowledgments}
This work is supported by the National Natural Science Foundation of China (Grants No. 12275109, No. 11834005, and No. 12247101) and the Supercomputing Center of Lanzhou University.

\bibliography{Ref}

\begin{thebibliography}{81}%
\makeatletter
\providecommand \@ifxundefined [1]{%
 \@ifx{#1\undefined}
}%
\providecommand \@ifnum [1]{%
 \ifnum #1\expandafter \@firstoftwo
 \else \expandafter \@secondoftwo
 \fi
}%
\providecommand \@ifx [1]{%
 \ifx #1\expandafter \@firstoftwo
 \else \expandafter \@secondoftwo
 \fi
}%
\providecommand \natexlab [1]{#1}%
\providecommand \enquote  [1]{``#1''}%
\providecommand \bibnamefont  [1]{#1}%
\providecommand \bibfnamefont [1]{#1}%
\providecommand \citenamefont [1]{#1}%
\providecommand \href@noop [0]{\@secondoftwo}%
\providecommand \href [0]{\begingroup \@sanitize@url \@href}%
\providecommand \@href[1]{\@@startlink{#1}\@@href}%
\providecommand \@@href[1]{\endgroup#1\@@endlink}%
\providecommand \@sanitize@url [0]{\catcode `\\12\catcode `\$12\catcode
  `\&12\catcode `\#12\catcode `\^12\catcode `\_12\catcode `\%12\relax}%
\providecommand \@@startlink[1]{}%
\providecommand \@@endlink[0]{}%
\providecommand \url  [0]{\begingroup\@sanitize@url \@url }%
\providecommand \@url [1]{\endgroup\@href {#1}{\urlprefix }}%
\providecommand \urlprefix  [0]{URL }%
\providecommand \Eprint [0]{\href }%
\providecommand \doibase [0]{https://doi.org/}%
\providecommand \selectlanguage [0]{\@gobble}%
\providecommand \bibinfo  [0]{\@secondoftwo}%
\providecommand \bibfield  [0]{\@secondoftwo}%
\providecommand \translation [1]{[#1]}%
\providecommand \BibitemOpen [0]{}%
\providecommand \bibitemStop [0]{}%
\providecommand \bibitemNoStop [0]{.\EOS\space}%
\providecommand \EOS [0]{\spacefactor3000\relax}%
\providecommand \BibitemShut  [1]{\csname bibitem#1\endcsname}%
\let\auto@bib@innerbib\@empty
\bibitem [{\citenamefont {Yuan}\ \emph {et~al.}(2022)\citenamefont {Yuan},
  \citenamefont {Cao}, \citenamefont {Kamra}, \citenamefont {Duine},\ and\
  \citenamefont {Yan}}]{YUAN20221}%
  \BibitemOpen
  \bibfield  {author} {\bibinfo {author} {\bibfnamefont {H.}~\bibnamefont
  {Yuan}}, \bibinfo {author} {\bibfnamefont {Y.}~\bibnamefont {Cao}}, \bibinfo
  {author} {\bibfnamefont {A.}~\bibnamefont {Kamra}}, \bibinfo {author}
  {\bibfnamefont {R.~A.}\ \bibnamefont {Duine}},\ and\ \bibinfo {author}
  {\bibfnamefont {P.}~\bibnamefont {Yan}},\ }\bibfield  {title} {\bibinfo
  {title} {Quantum magnonics: When magnon spintronics meets quantum information
  science},\ }\href
  {https://doi.org/https://doi.org/10.1016/j.physrep.2022.03.002} {\bibfield
  {journal} {\bibinfo  {journal} {Physics Reports}\ }\textbf {\bibinfo {volume}
  {965}},\ \bibinfo {pages} {1} (\bibinfo {year} {2022})}\BibitemShut {NoStop}%
\bibitem [{\citenamefont {Lachance-Quirion}\ \emph {et~al.}(2019)\citenamefont
  {Lachance-Quirion}, \citenamefont {Tabuchi}, \citenamefont {Gloppe},
  \citenamefont {Usami},\ and\ \citenamefont
  {Nakamura}}]{Lachance-Quirion_2019}%
  \BibitemOpen
  \bibfield  {author} {\bibinfo {author} {\bibfnamefont {D.}~\bibnamefont
  {Lachance-Quirion}}, \bibinfo {author} {\bibfnamefont {Y.}~\bibnamefont
  {Tabuchi}}, \bibinfo {author} {\bibfnamefont {A.}~\bibnamefont {Gloppe}},
  \bibinfo {author} {\bibfnamefont {K.}~\bibnamefont {Usami}},\ and\ \bibinfo
  {author} {\bibfnamefont {Y.}~\bibnamefont {Nakamura}},\ }\bibfield  {title}
  {\bibinfo {title} {Hybrid quantum systems based on magnonics},\ }\href
  {https://doi.org/10.7567/1882-0786/ab248d} {\bibfield  {journal} {\bibinfo
  {journal} {Applied Physics Express}\ }\textbf {\bibinfo {volume} {12}},\
  \bibinfo {pages} {070101} (\bibinfo {year} {2019})}\BibitemShut {NoStop}%
\bibitem [{\citenamefont {{Zare Rameshti}}\ \emph {et~al.}(2022)\citenamefont
  {{Zare Rameshti}}, \citenamefont {{Viola Kusminskiy}}, \citenamefont {Haigh},
  \citenamefont {Usami}, \citenamefont {Lachance-Quirion}, \citenamefont
  {Nakamura}, \citenamefont {Hu}, \citenamefont {Tang}, \citenamefont {Bauer},\
  and\ \citenamefont {Blanter}}]{ZARERAMESHTI20221}%
  \BibitemOpen
  \bibfield  {author} {\bibinfo {author} {\bibfnamefont {B.}~\bibnamefont
  {{Zare Rameshti}}}, \bibinfo {author} {\bibfnamefont {S.}~\bibnamefont
  {{Viola Kusminskiy}}}, \bibinfo {author} {\bibfnamefont {J.~A.}\ \bibnamefont
  {Haigh}}, \bibinfo {author} {\bibfnamefont {K.}~\bibnamefont {Usami}},
  \bibinfo {author} {\bibfnamefont {D.}~\bibnamefont {Lachance-Quirion}},
  \bibinfo {author} {\bibfnamefont {Y.}~\bibnamefont {Nakamura}}, \bibinfo
  {author} {\bibfnamefont {C.-M.}\ \bibnamefont {Hu}}, \bibinfo {author}
  {\bibfnamefont {H.~X.}\ \bibnamefont {Tang}}, \bibinfo {author}
  {\bibfnamefont {G.~E.}\ \bibnamefont {Bauer}},\ and\ \bibinfo {author}
  {\bibfnamefont {Y.~M.}\ \bibnamefont {Blanter}},\ }\bibfield  {title}
  {\bibinfo {title} {Cavity magnonics},\ }\href
  {https://doi.org/https://doi.org/10.1016/j.physrep.2022.06.001} {\bibfield
  {journal} {\bibinfo  {journal} {Physics Reports}\ }\textbf {\bibinfo {volume}
  {979}},\ \bibinfo {pages} {1} (\bibinfo {year} {2022})}\BibitemShut {NoStop}%
\bibitem [{\citenamefont {Zhang}\ \emph {et~al.}(2016)\citenamefont {Zhang},
  \citenamefont {Zou}, \citenamefont {Jiang},\ and\ \citenamefont
  {Tang}}]{doi:10.1126/sciadv.1501286}%
  \BibitemOpen
  \bibfield  {author} {\bibinfo {author} {\bibfnamefont {X.}~\bibnamefont
  {Zhang}}, \bibinfo {author} {\bibfnamefont {C.-L.}\ \bibnamefont {Zou}},
  \bibinfo {author} {\bibfnamefont {L.}~\bibnamefont {Jiang}},\ and\ \bibinfo
  {author} {\bibfnamefont {H.~X.}\ \bibnamefont {Tang}},\ }\bibfield  {title}
  {\bibinfo {title} {Cavity magnomechanics},\ }\href
  {https://doi.org/10.1126/sciadv.1501286} {\bibfield  {journal} {\bibinfo
  {journal} {Science Advances}\ }\textbf {\bibinfo {volume} {2}},\ \bibinfo
  {pages} {e1501286} (\bibinfo {year} {2016})}\BibitemShut {NoStop}%
\bibitem [{\citenamefont {Chumak}\ \emph {et~al.}(2014)\citenamefont {Chumak},
  \citenamefont {Serga},\ and\ \citenamefont
  {Hillebrands}}]{10.1038/ncomms5700}%
  \BibitemOpen
  \bibfield  {author} {\bibinfo {author} {\bibfnamefont {A.~V.}\ \bibnamefont
  {Chumak}}, \bibinfo {author} {\bibfnamefont {A.~A.}\ \bibnamefont {Serga}},\
  and\ \bibinfo {author} {\bibfnamefont {B.}~\bibnamefont {Hillebrands}},\
  }\bibfield  {title} {\bibinfo {title} {Magnon transistor for all-magnon data
  processing},\ }\href {https://doi.org/10.1038/ncomms5700} {\bibfield
  {journal} {\bibinfo  {journal} {Nature Communications}\ }\textbf {\bibinfo
  {volume} {5}},\ \bibinfo {pages} {4700} (\bibinfo {year} {2014})}\BibitemShut
  {NoStop}%
\bibitem [{\citenamefont {Nambu}\ \emph {et~al.}(2020)\citenamefont {Nambu},
  \citenamefont {Barker}, \citenamefont {Okino}, \citenamefont {Kikkawa},
  \citenamefont {Shiomi}, \citenamefont {Enderle}, \citenamefont {Weber},
  \citenamefont {Winn}, \citenamefont {Graves-Brook}, \citenamefont
  {Tranquada}, \citenamefont {Ziman}, \citenamefont {Fujita}, \citenamefont
  {Bauer}, \citenamefont {Saitoh},\ and\ \citenamefont
  {Kakurai}}]{PhysRevLett.125.027201}%
  \BibitemOpen
  \bibfield  {author} {\bibinfo {author} {\bibfnamefont {Y.}~\bibnamefont
  {Nambu}}, \bibinfo {author} {\bibfnamefont {J.}~\bibnamefont {Barker}},
  \bibinfo {author} {\bibfnamefont {Y.}~\bibnamefont {Okino}}, \bibinfo
  {author} {\bibfnamefont {T.}~\bibnamefont {Kikkawa}}, \bibinfo {author}
  {\bibfnamefont {Y.}~\bibnamefont {Shiomi}}, \bibinfo {author} {\bibfnamefont
  {M.}~\bibnamefont {Enderle}}, \bibinfo {author} {\bibfnamefont
  {T.}~\bibnamefont {Weber}}, \bibinfo {author} {\bibfnamefont
  {B.}~\bibnamefont {Winn}}, \bibinfo {author} {\bibfnamefont {M.}~\bibnamefont
  {Graves-Brook}}, \bibinfo {author} {\bibfnamefont {J.~M.}\ \bibnamefont
  {Tranquada}}, \bibinfo {author} {\bibfnamefont {T.}~\bibnamefont {Ziman}},
  \bibinfo {author} {\bibfnamefont {M.}~\bibnamefont {Fujita}}, \bibinfo
  {author} {\bibfnamefont {G.~E.~W.}\ \bibnamefont {Bauer}}, \bibinfo {author}
  {\bibfnamefont {E.}~\bibnamefont {Saitoh}},\ and\ \bibinfo {author}
  {\bibfnamefont {K.}~\bibnamefont {Kakurai}},\ }\bibfield  {title} {\bibinfo
  {title} {Observation of magnon polarization},\ }\href
  {https://doi.org/10.1103/PhysRevLett.125.027201} {\bibfield  {journal}
  {\bibinfo  {journal} {Phys. Rev. Lett.}\ }\textbf {\bibinfo {volume} {125}},\
  \bibinfo {pages} {027201} (\bibinfo {year} {2020})}\BibitemShut {NoStop}%
\bibitem [{\citenamefont {Sun}\ \emph {et~al.}(2021)\citenamefont {Sun},
  \citenamefont {Zheng}, \citenamefont {Xiao}, \citenamefont {Gong},
  \citenamefont {He},\ and\ \citenamefont {Xia}}]{PhysRevLett.127.087203}%
  \BibitemOpen
  \bibfield  {author} {\bibinfo {author} {\bibfnamefont {F.-X.}\ \bibnamefont
  {Sun}}, \bibinfo {author} {\bibfnamefont {S.-S.}\ \bibnamefont {Zheng}},
  \bibinfo {author} {\bibfnamefont {Y.}~\bibnamefont {Xiao}}, \bibinfo {author}
  {\bibfnamefont {Q.}~\bibnamefont {Gong}}, \bibinfo {author} {\bibfnamefont
  {Q.}~\bibnamefont {He}},\ and\ \bibinfo {author} {\bibfnamefont
  {K.}~\bibnamefont {Xia}},\ }\bibfield  {title} {\bibinfo {title} {Remote
  generation of magnon {S}chr\"odinger cat state via magnon-photon
  entanglement},\ }\href {https://doi.org/10.1103/PhysRevLett.127.087203}
  {\bibfield  {journal} {\bibinfo  {journal} {Phys. Rev. Lett.}\ }\textbf
  {\bibinfo {volume} {127}},\ \bibinfo {pages} {087203} (\bibinfo {year}
  {2021})}\BibitemShut {NoStop}%
\bibitem [{\citenamefont {Yuan}\ \emph {et~al.}(2020)\citenamefont {Yuan},
  \citenamefont {Yan}, \citenamefont {Zheng}, \citenamefont {He}, \citenamefont
  {Xia},\ and\ \citenamefont {Yung}}]{PhysRevLett.124.053602}%
  \BibitemOpen
  \bibfield  {author} {\bibinfo {author} {\bibfnamefont {H.~Y.}\ \bibnamefont
  {Yuan}}, \bibinfo {author} {\bibfnamefont {P.}~\bibnamefont {Yan}}, \bibinfo
  {author} {\bibfnamefont {S.}~\bibnamefont {Zheng}}, \bibinfo {author}
  {\bibfnamefont {Q.~Y.}\ \bibnamefont {He}}, \bibinfo {author} {\bibfnamefont
  {K.}~\bibnamefont {Xia}},\ and\ \bibinfo {author} {\bibfnamefont {M.-H.}\
  \bibnamefont {Yung}},\ }\bibfield  {title} {\bibinfo {title} {Steady {B}ell
  state generation via magnon-photon coupling},\ }\href
  {https://doi.org/10.1103/PhysRevLett.124.053602} {\bibfield  {journal}
  {\bibinfo  {journal} {Phys. Rev. Lett.}\ }\textbf {\bibinfo {volume} {124}},\
  \bibinfo {pages} {053602} (\bibinfo {year} {2020})}\BibitemShut {NoStop}%
\bibitem [{\citenamefont {Prokop}\ \emph {et~al.}(2009)\citenamefont {Prokop},
  \citenamefont {Tang}, \citenamefont {Zhang}, \citenamefont {Tudosa},
  \citenamefont {Peixoto}, \citenamefont {Zakeri},\ and\ \citenamefont
  {Kirschner}}]{PhysRevLett.102.177206}%
  \BibitemOpen
  \bibfield  {author} {\bibinfo {author} {\bibfnamefont {J.}~\bibnamefont
  {Prokop}}, \bibinfo {author} {\bibfnamefont {W.~X.}\ \bibnamefont {Tang}},
  \bibinfo {author} {\bibfnamefont {Y.}~\bibnamefont {Zhang}}, \bibinfo
  {author} {\bibfnamefont {I.}~\bibnamefont {Tudosa}}, \bibinfo {author}
  {\bibfnamefont {T.~R.~F.}\ \bibnamefont {Peixoto}}, \bibinfo {author}
  {\bibfnamefont {K.}~\bibnamefont {Zakeri}},\ and\ \bibinfo {author}
  {\bibfnamefont {J.}~\bibnamefont {Kirschner}},\ }\bibfield  {title} {\bibinfo
  {title} {Magnons in a ferromagnetic monolayer},\ }\href
  {https://doi.org/10.1103/PhysRevLett.102.177206} {\bibfield  {journal}
  {\bibinfo  {journal} {Phys. Rev. Lett.}\ }\textbf {\bibinfo {volume} {102}},\
  \bibinfo {pages} {177206} (\bibinfo {year} {2009})}\BibitemShut {NoStop}%
\bibitem [{\citenamefont {Wang}\ \emph {et~al.}(2022)\citenamefont {Wang},
  \citenamefont {Wang}, \citenamefont {Yao}, \citenamefont {Shen},
  \citenamefont {Wu}, \citenamefont {Qian}, \citenamefont {Li}, \citenamefont
  {Zhu},\ and\ \citenamefont {You}}]{10.1038/s41467-022-35174-9}%
  \BibitemOpen
  \bibfield  {author} {\bibinfo {author} {\bibfnamefont {Z.-Q.}\ \bibnamefont
  {Wang}}, \bibinfo {author} {\bibfnamefont {Y.-P.}\ \bibnamefont {Wang}},
  \bibinfo {author} {\bibfnamefont {J.}~\bibnamefont {Yao}}, \bibinfo {author}
  {\bibfnamefont {R.-C.}\ \bibnamefont {Shen}}, \bibinfo {author}
  {\bibfnamefont {W.-J.}\ \bibnamefont {Wu}}, \bibinfo {author} {\bibfnamefont
  {J.}~\bibnamefont {Qian}}, \bibinfo {author} {\bibfnamefont {J.}~\bibnamefont
  {Li}}, \bibinfo {author} {\bibfnamefont {S.-Y.}\ \bibnamefont {Zhu}},\ and\
  \bibinfo {author} {\bibfnamefont {J.~Q.}\ \bibnamefont {You}},\ }\bibfield
  {title} {\bibinfo {title} {Giant spin ensembles in waveguide magnonics},\
  }\href {https://doi.org/10.1038/s41467-022-35174-9} {\bibfield  {journal}
  {\bibinfo  {journal} {Nature Communications}\ }\textbf {\bibinfo {volume}
  {13}},\ \bibinfo {pages} {7580} (\bibinfo {year} {2022})}\BibitemShut
  {NoStop}%
\bibitem [{\citenamefont {Li}\ \emph {et~al.}(2022{\natexlab{a}})\citenamefont
  {Li}, \citenamefont {Ma}, \citenamefont {Chen}, \citenamefont {Xie},\ and\
  \citenamefont {Ma}}]{10.1063/5.0121314}%
  \BibitemOpen
  \bibfield  {author} {\bibinfo {author} {\bibfnamefont {Z.}~\bibnamefont
  {Li}}, \bibinfo {author} {\bibfnamefont {M.}~\bibnamefont {Ma}}, \bibinfo
  {author} {\bibfnamefont {Z.}~\bibnamefont {Chen}}, \bibinfo {author}
  {\bibfnamefont {K.}~\bibnamefont {Xie}},\ and\ \bibinfo {author}
  {\bibfnamefont {F.}~\bibnamefont {Ma}},\ }\bibfield  {title} {\bibinfo
  {title} {{Interaction between magnon and skyrmion: Toward quantum
  magnonics}},\ }\href {https://doi.org/10.1063/5.0121314} {\bibfield
  {journal} {\bibinfo  {journal} {Journal of Applied Physics}\ }\textbf
  {\bibinfo {volume} {132}},\ \bibinfo {pages} {210702} (\bibinfo {year}
  {2022}{\natexlab{a}})}\BibitemShut {NoStop}%
\bibitem [{\citenamefont {Hisatomi}\ \emph {et~al.}(2016)\citenamefont
  {Hisatomi}, \citenamefont {Osada}, \citenamefont {Tabuchi}, \citenamefont
  {Ishikawa}, \citenamefont {Noguchi}, \citenamefont {Yamazaki}, \citenamefont
  {Usami},\ and\ \citenamefont {Nakamura}}]{PhysRevB.93.174427}%
  \BibitemOpen
  \bibfield  {author} {\bibinfo {author} {\bibfnamefont {R.}~\bibnamefont
  {Hisatomi}}, \bibinfo {author} {\bibfnamefont {A.}~\bibnamefont {Osada}},
  \bibinfo {author} {\bibfnamefont {Y.}~\bibnamefont {Tabuchi}}, \bibinfo
  {author} {\bibfnamefont {T.}~\bibnamefont {Ishikawa}}, \bibinfo {author}
  {\bibfnamefont {A.}~\bibnamefont {Noguchi}}, \bibinfo {author} {\bibfnamefont
  {R.}~\bibnamefont {Yamazaki}}, \bibinfo {author} {\bibfnamefont
  {K.}~\bibnamefont {Usami}},\ and\ \bibinfo {author} {\bibfnamefont
  {Y.}~\bibnamefont {Nakamura}},\ }\bibfield  {title} {\bibinfo {title}
  {Bidirectional conversion between microwave and light via ferromagnetic
  magnons},\ }\href {https://doi.org/10.1103/PhysRevB.93.174427} {\bibfield
  {journal} {\bibinfo  {journal} {Phys. Rev. B}\ }\textbf {\bibinfo {volume}
  {93}},\ \bibinfo {pages} {174427} (\bibinfo {year} {2016})}\BibitemShut
  {NoStop}%
\bibitem [{\citenamefont {Zhang}\ \emph {et~al.}(2015)\citenamefont {Zhang},
  \citenamefont {Zou}, \citenamefont {Zhu}, \citenamefont {Marquardt},
  \citenamefont {Jiang},\ and\ \citenamefont {Tang}}]{Zhang2015}%
  \BibitemOpen
  \bibfield  {author} {\bibinfo {author} {\bibfnamefont {X.}~\bibnamefont
  {Zhang}}, \bibinfo {author} {\bibfnamefont {C.-L.}\ \bibnamefont {Zou}},
  \bibinfo {author} {\bibfnamefont {N.}~\bibnamefont {Zhu}}, \bibinfo {author}
  {\bibfnamefont {F.}~\bibnamefont {Marquardt}}, \bibinfo {author}
  {\bibfnamefont {L.}~\bibnamefont {Jiang}},\ and\ \bibinfo {author}
  {\bibfnamefont {H.~X.}\ \bibnamefont {Tang}},\ }\bibfield  {title} {\bibinfo
  {title} {Magnon dark modes and gradient memory},\ }\href
  {https://doi.org/10.1038/ncomms9914} {\bibfield  {journal} {\bibinfo
  {journal} {Nature Communications}\ }\textbf {\bibinfo {volume} {6}},\
  \bibinfo {pages} {8914} (\bibinfo {year} {2015})}\BibitemShut {NoStop}%
\bibitem [{\citenamefont {Potts}\ \emph {et~al.}(2020)\citenamefont {Potts},
  \citenamefont {Bittencourt}, \citenamefont {Kusminskiy},\ and\ \citenamefont
  {Davis}}]{PhysRevApplied.13.064001}%
  \BibitemOpen
  \bibfield  {author} {\bibinfo {author} {\bibfnamefont {C.}~\bibnamefont
  {Potts}}, \bibinfo {author} {\bibfnamefont {V.}~\bibnamefont {Bittencourt}},
  \bibinfo {author} {\bibfnamefont {S.~V.}\ \bibnamefont {Kusminskiy}},\ and\
  \bibinfo {author} {\bibfnamefont {J.}~\bibnamefont {Davis}},\ }\bibfield
  {title} {\bibinfo {title} {Magnon-phonon quantum correlation thermometry},\
  }\href {https://doi.org/10.1103/PhysRevApplied.13.064001} {\bibfield
  {journal} {\bibinfo  {journal} {Phys. Rev. Appl.}\ }\textbf {\bibinfo
  {volume} {13}},\ \bibinfo {pages} {064001} (\bibinfo {year}
  {2020})}\BibitemShut {NoStop}%
\bibitem [{\citenamefont {Crescini}\ \emph {et~al.}(2020)\citenamefont
  {Crescini}, \citenamefont {Braggio}, \citenamefont {Carugno}, \citenamefont
  {Ortolan},\ and\ \citenamefont {Ruoso}}]{Crescini2020}%
  \BibitemOpen
  \bibfield  {author} {\bibinfo {author} {\bibfnamefont {N.}~\bibnamefont
  {Crescini}}, \bibinfo {author} {\bibfnamefont {C.}~\bibnamefont {Braggio}},
  \bibinfo {author} {\bibfnamefont {G.}~\bibnamefont {Carugno}}, \bibinfo
  {author} {\bibfnamefont {A.}~\bibnamefont {Ortolan}},\ and\ \bibinfo {author}
  {\bibfnamefont {G.}~\bibnamefont {Ruoso}},\ }\bibfield  {title} {\bibinfo
  {title} {Cavity magnon polariton based precision magnetometry},\ }\href
  {https://doi.org/10.1063/5.0024369} {\bibfield  {journal} {\bibinfo
  {journal} {Applied Physics Letters}\ }\textbf {\bibinfo {volume} {117}},\
  \bibinfo {pages} {144001} (\bibinfo {year} {2020})}\BibitemShut {NoStop}%
\bibitem [{\citenamefont {Wang}\ \emph {et~al.}(2019)\citenamefont {Wang},
  \citenamefont {Rao}, \citenamefont {Yang}, \citenamefont {Xu}, \citenamefont
  {Gui}, \citenamefont {Yao}, \citenamefont {You},\ and\ \citenamefont
  {Hu}}]{PhysRevLett.123.127202}%
  \BibitemOpen
  \bibfield  {author} {\bibinfo {author} {\bibfnamefont {Y.-P.}\ \bibnamefont
  {Wang}}, \bibinfo {author} {\bibfnamefont {J.~W.}\ \bibnamefont {Rao}},
  \bibinfo {author} {\bibfnamefont {Y.}~\bibnamefont {Yang}}, \bibinfo {author}
  {\bibfnamefont {P.-C.}\ \bibnamefont {Xu}}, \bibinfo {author} {\bibfnamefont
  {Y.~S.}\ \bibnamefont {Gui}}, \bibinfo {author} {\bibfnamefont {B.~M.}\
  \bibnamefont {Yao}}, \bibinfo {author} {\bibfnamefont {J.~Q.}\ \bibnamefont
  {You}},\ and\ \bibinfo {author} {\bibfnamefont {C.-M.}\ \bibnamefont {Hu}},\
  }\bibfield  {title} {\bibinfo {title} {Nonreciprocity and unidirectional
  invisibility in cavity magnonics},\ }\href
  {https://doi.org/10.1103/PhysRevLett.123.127202} {\bibfield  {journal}
  {\bibinfo  {journal} {Phys. Rev. Lett.}\ }\textbf {\bibinfo {volume} {123}},\
  \bibinfo {pages} {127202} (\bibinfo {year} {2019})}\BibitemShut {NoStop}%
\bibitem [{\citenamefont {Rusconi}\ \emph {et~al.}(2019)\citenamefont
  {Rusconi}, \citenamefont {Schuetz}, \citenamefont {Gieseler}, \citenamefont
  {Lukin},\ and\ \citenamefont {Romero-Isart}}]{PhysRevA.100.022343}%
  \BibitemOpen
  \bibfield  {author} {\bibinfo {author} {\bibfnamefont {C.~C.}\ \bibnamefont
  {Rusconi}}, \bibinfo {author} {\bibfnamefont {M.~J.~A.}\ \bibnamefont
  {Schuetz}}, \bibinfo {author} {\bibfnamefont {J.}~\bibnamefont {Gieseler}},
  \bibinfo {author} {\bibfnamefont {M.~D.}\ \bibnamefont {Lukin}},\ and\
  \bibinfo {author} {\bibfnamefont {O.}~\bibnamefont {Romero-Isart}},\
  }\bibfield  {title} {\bibinfo {title} {Hybrid architecture for engineering
  magnonic quantum networks},\ }\href
  {https://doi.org/10.1103/PhysRevA.100.022343} {\bibfield  {journal} {\bibinfo
   {journal} {Phys. Rev. A}\ }\textbf {\bibinfo {volume} {100}},\ \bibinfo
  {pages} {022343} (\bibinfo {year} {2019})}\BibitemShut {NoStop}%
\bibitem [{\citenamefont {Lachance-Quirion}\ \emph {et~al.}(2020)\citenamefont
  {Lachance-Quirion}, \citenamefont {Wolski}, \citenamefont {Tabuchi},
  \citenamefont {Kono}, \citenamefont {Usami},\ and\ \citenamefont
  {Nakamura}}]{doi:10.1126/science.aaz9236}%
  \BibitemOpen
  \bibfield  {author} {\bibinfo {author} {\bibfnamefont {D.}~\bibnamefont
  {Lachance-Quirion}}, \bibinfo {author} {\bibfnamefont {S.~P.}\ \bibnamefont
  {Wolski}}, \bibinfo {author} {\bibfnamefont {Y.}~\bibnamefont {Tabuchi}},
  \bibinfo {author} {\bibfnamefont {S.}~\bibnamefont {Kono}}, \bibinfo {author}
  {\bibfnamefont {K.}~\bibnamefont {Usami}},\ and\ \bibinfo {author}
  {\bibfnamefont {Y.}~\bibnamefont {Nakamura}},\ }\bibfield  {title} {\bibinfo
  {title} {Entanglement-based single-shot detection of a single magnon with a
  superconducting qubit},\ }\href {https://doi.org/10.1126/science.aaz9236}
  {\bibfield  {journal} {\bibinfo  {journal} {Science}\ }\textbf {\bibinfo
  {volume} {367}},\ \bibinfo {pages} {425} (\bibinfo {year}
  {2020})}\BibitemShut {NoStop}%
\bibitem [{\citenamefont {Solanki}\ \emph {et~al.}(2022)\citenamefont
  {Solanki}, \citenamefont {Bogdanov}, \citenamefont {Rahman}, \citenamefont
  {Rustagi}, \citenamefont {Dilley}, \citenamefont {Shen}, \citenamefont
  {Tong}, \citenamefont {Debashis}, \citenamefont {Chen}, \citenamefont
  {Appenzeller}, \citenamefont {Chen}, \citenamefont {Shalaev},\ and\
  \citenamefont {Upadhyaya}}]{PhysRevResearch.4.L012025}%
  \BibitemOpen
  \bibfield  {author} {\bibinfo {author} {\bibfnamefont {A.~B.}\ \bibnamefont
  {Solanki}}, \bibinfo {author} {\bibfnamefont {S.~I.}\ \bibnamefont
  {Bogdanov}}, \bibinfo {author} {\bibfnamefont {M.~M.}\ \bibnamefont
  {Rahman}}, \bibinfo {author} {\bibfnamefont {A.}~\bibnamefont {Rustagi}},
  \bibinfo {author} {\bibfnamefont {N.~R.}\ \bibnamefont {Dilley}}, \bibinfo
  {author} {\bibfnamefont {T.}~\bibnamefont {Shen}}, \bibinfo {author}
  {\bibfnamefont {W.}~\bibnamefont {Tong}}, \bibinfo {author} {\bibfnamefont
  {P.}~\bibnamefont {Debashis}}, \bibinfo {author} {\bibfnamefont
  {Z.}~\bibnamefont {Chen}}, \bibinfo {author} {\bibfnamefont {J.}~\bibnamefont
  {Appenzeller}}, \bibinfo {author} {\bibfnamefont {Y.~P.}\ \bibnamefont
  {Chen}}, \bibinfo {author} {\bibfnamefont {V.~M.}\ \bibnamefont {Shalaev}},\
  and\ \bibinfo {author} {\bibfnamefont {P.}~\bibnamefont {Upadhyaya}},\
  }\bibfield  {title} {\bibinfo {title} {Electric field control of interaction
  between magnons and quantum spin defects},\ }\href
  {https://doi.org/10.1103/PhysRevResearch.4.L012025} {\bibfield  {journal}
  {\bibinfo  {journal} {Phys. Rev. Research}\ }\textbf {\bibinfo {volume}
  {4}},\ \bibinfo {pages} {L012025} (\bibinfo {year} {2022})}\BibitemShut
  {NoStop}%
\bibitem [{\citenamefont {Prananto}\ \emph {et~al.}(2021)\citenamefont
  {Prananto}, \citenamefont {Kainuma}, \citenamefont {Hayashi}, \citenamefont
  {Mizuochi}, \citenamefont {Uchida},\ and\ \citenamefont
  {An}}]{PhysRevApplied.16.064058}%
  \BibitemOpen
  \bibfield  {author} {\bibinfo {author} {\bibfnamefont {D.}~\bibnamefont
  {Prananto}}, \bibinfo {author} {\bibfnamefont {Y.}~\bibnamefont {Kainuma}},
  \bibinfo {author} {\bibfnamefont {K.}~\bibnamefont {Hayashi}}, \bibinfo
  {author} {\bibfnamefont {N.}~\bibnamefont {Mizuochi}}, \bibinfo {author}
  {\bibfnamefont {K.-i.}\ \bibnamefont {Uchida}},\ and\ \bibinfo {author}
  {\bibfnamefont {T.}~\bibnamefont {An}},\ }\bibfield  {title} {\bibinfo
  {title} {Probing thermal magnon current mediated by coherent magnon via
  nitrogen-vacancy centers in diamond},\ }\href
  {https://doi.org/10.1103/PhysRevApplied.16.064058} {\bibfield  {journal}
  {\bibinfo  {journal} {Phys. Rev. Appl.}\ }\textbf {\bibinfo {volume} {16}},\
  \bibinfo {pages} {064058} (\bibinfo {year} {2021})}\BibitemShut {NoStop}%
\bibitem [{\citenamefont {Tabuchi}\ \emph {et~al.}(2015)\citenamefont
  {Tabuchi}, \citenamefont {Ishino}, \citenamefont {Noguchi}, \citenamefont
  {Ishikawa}, \citenamefont {Yamazaki}, \citenamefont {Usami},\ and\
  \citenamefont {Nakamura}}]{doi:10.1126/science.aaa3693}%
  \BibitemOpen
  \bibfield  {author} {\bibinfo {author} {\bibfnamefont {Y.}~\bibnamefont
  {Tabuchi}}, \bibinfo {author} {\bibfnamefont {S.}~\bibnamefont {Ishino}},
  \bibinfo {author} {\bibfnamefont {A.}~\bibnamefont {Noguchi}}, \bibinfo
  {author} {\bibfnamefont {T.}~\bibnamefont {Ishikawa}}, \bibinfo {author}
  {\bibfnamefont {R.}~\bibnamefont {Yamazaki}}, \bibinfo {author}
  {\bibfnamefont {K.}~\bibnamefont {Usami}},\ and\ \bibinfo {author}
  {\bibfnamefont {Y.}~\bibnamefont {Nakamura}},\ }\bibfield  {title} {\bibinfo
  {title} {Coherent coupling between a ferromagnetic magnon and a
  superconducting qubit},\ }\href {https://doi.org/10.1126/science.aaa3693}
  {\bibfield  {journal} {\bibinfo  {journal} {Science}\ }\textbf {\bibinfo
  {volume} {349}},\ \bibinfo {pages} {405} (\bibinfo {year}
  {2015})}\BibitemShut {NoStop}%
\bibitem [{\citenamefont {Neuman}\ \emph {et~al.}(2020)\citenamefont {Neuman},
  \citenamefont {Wang},\ and\ \citenamefont {Narang}}]{PhysRevLett.125.247702}%
  \BibitemOpen
  \bibfield  {author} {\bibinfo {author} {\bibfnamefont {T.}~\bibnamefont
  {Neuman}}, \bibinfo {author} {\bibfnamefont {D.~S.}\ \bibnamefont {Wang}},\
  and\ \bibinfo {author} {\bibfnamefont {P.}~\bibnamefont {Narang}},\
  }\bibfield  {title} {\bibinfo {title} {Nanomagnonic cavities for strong
  spin-magnon coupling and magnon-mediated spin-spin interactions},\ }\href
  {https://doi.org/10.1103/PhysRevLett.125.247702} {\bibfield  {journal}
  {\bibinfo  {journal} {Phys. Rev. Lett.}\ }\textbf {\bibinfo {volume} {125}},\
  \bibinfo {pages} {247702} (\bibinfo {year} {2020})}\BibitemShut {NoStop}%
\bibitem [{\citenamefont {Fukami}\ \emph {et~al.}(2021)\citenamefont {Fukami},
  \citenamefont {Candido}, \citenamefont {Awschalom},\ and\ \citenamefont
  {Flatt\'e}}]{PRXQuantum.2.040314}%
  \BibitemOpen
  \bibfield  {author} {\bibinfo {author} {\bibfnamefont {M.}~\bibnamefont
  {Fukami}}, \bibinfo {author} {\bibfnamefont {D.~R.}\ \bibnamefont {Candido}},
  \bibinfo {author} {\bibfnamefont {D.~D.}\ \bibnamefont {Awschalom}},\ and\
  \bibinfo {author} {\bibfnamefont {M.~E.}\ \bibnamefont {Flatt\'e}},\
  }\bibfield  {title} {\bibinfo {title} {Opportunities for long-range
  magnon-mediated entanglement of spin qubits via on- and off-resonant
  coupling},\ }\href {https://doi.org/10.1103/PRXQuantum.2.040314} {\bibfield
  {journal} {\bibinfo  {journal} {PRX Quantum}\ }\textbf {\bibinfo {volume}
  {2}},\ \bibinfo {pages} {040314} (\bibinfo {year} {2021})}\BibitemShut
  {NoStop}%
\bibitem [{\citenamefont {Gonzalez-Ballestero}\ \emph
  {et~al.}(2022)\citenamefont {Gonzalez-Ballestero}, \citenamefont {van~der
  Sar},\ and\ \citenamefont {Romero-Isart}}]{PhysRevB.105.075410}%
  \BibitemOpen
  \bibfield  {author} {\bibinfo {author} {\bibfnamefont {C.}~\bibnamefont
  {Gonzalez-Ballestero}}, \bibinfo {author} {\bibfnamefont {T.}~\bibnamefont
  {van~der Sar}},\ and\ \bibinfo {author} {\bibfnamefont {O.}~\bibnamefont
  {Romero-Isart}},\ }\bibfield  {title} {\bibinfo {title} {Towards a quantum
  interface between spin waves and paramagnetic spin baths},\ }\href
  {https://doi.org/10.1103/PhysRevB.105.075410} {\bibfield  {journal} {\bibinfo
   {journal} {Phys. Rev. B}\ }\textbf {\bibinfo {volume} {105}},\ \bibinfo
  {pages} {075410} (\bibinfo {year} {2022})}\BibitemShut {NoStop}%
\bibitem [{\citenamefont {Ullah}\ \emph {et~al.}(2022)\citenamefont {Ullah},
  \citenamefont {K\"ose}, \citenamefont {Yagan}, \citenamefont {Onba\ifmmode
  \mbox{\c{s}}\else \c{s}\fi{}l\ifmmode \imath \else~\i \fi{}},\ and\
  \citenamefont {M\"ustecapl\ifmmode \imath \else \i
  \fi{}o\ifmmode~\breve{g}\else \u{g}\fi{}lu}}]{PhysRevResearch.4.023221}%
  \BibitemOpen
  \bibfield  {author} {\bibinfo {author} {\bibfnamefont {K.}~\bibnamefont
  {Ullah}}, \bibinfo {author} {\bibfnamefont {E.}~\bibnamefont {K\"ose}},
  \bibinfo {author} {\bibfnamefont {R.}~\bibnamefont {Yagan}}, \bibinfo
  {author} {\bibfnamefont {M.~C.}\ \bibnamefont {Onba\ifmmode \mbox{\c{s}}\else
  \c{s}\fi{}l\ifmmode \imath \else~\i \fi{}}},\ and\ \bibinfo {author}
  {\bibfnamefont {O.~E.}\ \bibnamefont {M\"ustecapl\ifmmode \imath \else \i
  \fi{}o\ifmmode~\breve{g}\else \u{g}\fi{}lu}},\ }\bibfield  {title} {\bibinfo
  {title} {Steady state entanglement of distant nitrogen-vacancy centers in a
  coherent thermal magnon bath},\ }\href
  {https://doi.org/10.1103/PhysRevResearch.4.023221} {\bibfield  {journal}
  {\bibinfo  {journal} {Phys. Rev. Research}\ }\textbf {\bibinfo {volume}
  {4}},\ \bibinfo {pages} {023221} (\bibinfo {year} {2022})}\BibitemShut
  {NoStop}%
\bibitem [{\citenamefont {Karanikolas}\ \emph {et~al.}(2022)\citenamefont
  {Karanikolas}, \citenamefont {Kuroda},\ and\ \citenamefont
  {Inoue}}]{PhysRevResearch.4.043180}%
  \BibitemOpen
  \bibfield  {author} {\bibinfo {author} {\bibfnamefont {V.}~\bibnamefont
  {Karanikolas}}, \bibinfo {author} {\bibfnamefont {T.}~\bibnamefont
  {Kuroda}},\ and\ \bibinfo {author} {\bibfnamefont {J.-i.}\ \bibnamefont
  {Inoue}},\ }\bibfield  {title} {\bibinfo {title} {Magnon-mediated spin
  entanglement in the strong-coupling regime},\ }\href
  {https://doi.org/10.1103/PhysRevResearch.4.043180} {\bibfield  {journal}
  {\bibinfo  {journal} {Phys. Rev. Res.}\ }\textbf {\bibinfo {volume} {4}},\
  \bibinfo {pages} {043180} (\bibinfo {year} {2022})}\BibitemShut {NoStop}%
\bibitem [{\citenamefont {An}\ \emph {et~al.}(2022)\citenamefont {An},
  \citenamefont {Kohno}, \citenamefont {Litvinenko}, \citenamefont {Seeger},
  \citenamefont {Naletov}, \citenamefont {Vila}, \citenamefont {de~Loubens},
  \citenamefont {Ben~Youssef}, \citenamefont {Vukadinovic}, \citenamefont
  {Bauer}, \citenamefont {Slavin}, \citenamefont {Tiberkevich},\ and\
  \citenamefont {Klein}}]{PhysRevX.12.011060}%
  \BibitemOpen
  \bibfield  {author} {\bibinfo {author} {\bibfnamefont {K.}~\bibnamefont
  {An}}, \bibinfo {author} {\bibfnamefont {R.}~\bibnamefont {Kohno}}, \bibinfo
  {author} {\bibfnamefont {A.~N.}\ \bibnamefont {Litvinenko}}, \bibinfo
  {author} {\bibfnamefont {R.~L.}\ \bibnamefont {Seeger}}, \bibinfo {author}
  {\bibfnamefont {V.~V.}\ \bibnamefont {Naletov}}, \bibinfo {author}
  {\bibfnamefont {L.}~\bibnamefont {Vila}}, \bibinfo {author} {\bibfnamefont
  {G.}~\bibnamefont {de~Loubens}}, \bibinfo {author} {\bibfnamefont
  {J.}~\bibnamefont {Ben~Youssef}}, \bibinfo {author} {\bibfnamefont
  {N.}~\bibnamefont {Vukadinovic}}, \bibinfo {author} {\bibfnamefont
  {G.~E.~W.}\ \bibnamefont {Bauer}}, \bibinfo {author} {\bibfnamefont {A.~N.}\
  \bibnamefont {Slavin}}, \bibinfo {author} {\bibfnamefont {V.~S.}\
  \bibnamefont {Tiberkevich}},\ and\ \bibinfo {author} {\bibfnamefont
  {O.}~\bibnamefont {Klein}},\ }\bibfield  {title} {\bibinfo {title} {Bright
  and dark states of two distant macrospins strongly coupled by phonons},\
  }\href {https://doi.org/10.1103/PhysRevX.12.011060} {\bibfield  {journal}
  {\bibinfo  {journal} {Phys. Rev. X}\ }\textbf {\bibinfo {volume} {12}},\
  \bibinfo {pages} {011060} (\bibinfo {year} {2022})}\BibitemShut {NoStop}%
\bibitem [{\citenamefont {Kounalakis}\ \emph {et~al.}(2022)\citenamefont
  {Kounalakis}, \citenamefont {Bauer},\ and\ \citenamefont
  {Blanter}}]{PhysRevLett.129.037205}%
  \BibitemOpen
  \bibfield  {author} {\bibinfo {author} {\bibfnamefont {M.}~\bibnamefont
  {Kounalakis}}, \bibinfo {author} {\bibfnamefont {G.~E.~W.}\ \bibnamefont
  {Bauer}},\ and\ \bibinfo {author} {\bibfnamefont {Y.~M.}\ \bibnamefont
  {Blanter}},\ }\bibfield  {title} {\bibinfo {title} {Analog quantum control of
  magnonic cat states on a chip by a superconducting qubit},\ }\href
  {https://doi.org/10.1103/PhysRevLett.129.037205} {\bibfield  {journal}
  {\bibinfo  {journal} {Phys. Rev. Lett.}\ }\textbf {\bibinfo {volume} {129}},\
  \bibinfo {pages} {037205} (\bibinfo {year} {2022})}\BibitemShut {NoStop}%
\bibitem [{\citenamefont {Wu}\ \emph {et~al.}(2021)\citenamefont {Wu},
  \citenamefont {Zhong}, \citenamefont {Cheng},\ and\ \citenamefont
  {Chen}}]{PhysRevA.103.052411}%
  \BibitemOpen
  \bibfield  {author} {\bibinfo {author} {\bibfnamefont {K.}~\bibnamefont
  {Wu}}, \bibinfo {author} {\bibfnamefont {W.-x.}\ \bibnamefont {Zhong}},
  \bibinfo {author} {\bibfnamefont {G.-l.}\ \bibnamefont {Cheng}},\ and\
  \bibinfo {author} {\bibfnamefont {A.-x.}\ \bibnamefont {Chen}},\ }\bibfield
  {title} {\bibinfo {title} {Phase-controlled multimagnon blockade and
  magnon-induced tunneling in a hybrid superconducting system},\ }\href
  {https://doi.org/10.1103/PhysRevA.103.052411} {\bibfield  {journal} {\bibinfo
   {journal} {Phys. Rev. A}\ }\textbf {\bibinfo {volume} {103}},\ \bibinfo
  {pages} {052411} (\bibinfo {year} {2021})}\BibitemShut {NoStop}%
\bibitem [{\citenamefont {Liu}\ \emph {et~al.}(2021)\citenamefont {Liu},
  \citenamefont {Xiong}, \citenamefont {Wu},\ and\ \citenamefont
  {Li}}]{PhysRevA.103.063702}%
  \BibitemOpen
  \bibfield  {author} {\bibinfo {author} {\bibfnamefont {Z.-X.}\ \bibnamefont
  {Liu}}, \bibinfo {author} {\bibfnamefont {H.}~\bibnamefont {Xiong}}, \bibinfo
  {author} {\bibfnamefont {M.-Y.}\ \bibnamefont {Wu}},\ and\ \bibinfo {author}
  {\bibfnamefont {Y.-q.}\ \bibnamefont {Li}},\ }\bibfield  {title} {\bibinfo
  {title} {Absorption of magnons in dispersively coupled hybrid quantum
  systems},\ }\href {https://doi.org/10.1103/PhysRevA.103.063702} {\bibfield
  {journal} {\bibinfo  {journal} {Phys. Rev. A}\ }\textbf {\bibinfo {volume}
  {103}},\ \bibinfo {pages} {063702} (\bibinfo {year} {2021})}\BibitemShut
  {NoStop}%
\bibitem [{\citenamefont {Xiong}\ \emph {et~al.}(2023)\citenamefont {Xiong},
  \citenamefont {Wang}, \citenamefont {Zhang},\ and\ \citenamefont
  {Chen}}]{PhysRevA.107.033516}%
  \BibitemOpen
  \bibfield  {author} {\bibinfo {author} {\bibfnamefont {W.}~\bibnamefont
  {Xiong}}, \bibinfo {author} {\bibfnamefont {M.}~\bibnamefont {Wang}},
  \bibinfo {author} {\bibfnamefont {G.-Q.}\ \bibnamefont {Zhang}},\ and\
  \bibinfo {author} {\bibfnamefont {J.}~\bibnamefont {Chen}},\ }\bibfield
  {title} {\bibinfo {title} {Optomechanical-interface-induced strong
  spin-magnon coupling},\ }\href {https://doi.org/10.1103/PhysRevA.107.033516}
  {\bibfield  {journal} {\bibinfo  {journal} {Phys. Rev. A}\ }\textbf {\bibinfo
  {volume} {107}},\ \bibinfo {pages} {033516} (\bibinfo {year}
  {2023})}\BibitemShut {NoStop}%
\bibitem [{\citenamefont {Hei}\ \emph {et~al.}(2023)\citenamefont {Hei},
  \citenamefont {Li}, \citenamefont {Pan},\ and\ \citenamefont
  {Nori}}]{PhysRevLett.130.073602}%
  \BibitemOpen
  \bibfield  {author} {\bibinfo {author} {\bibfnamefont {X.-L.}\ \bibnamefont
  {Hei}}, \bibinfo {author} {\bibfnamefont {P.-B.}\ \bibnamefont {Li}},
  \bibinfo {author} {\bibfnamefont {X.-F.}\ \bibnamefont {Pan}},\ and\ \bibinfo
  {author} {\bibfnamefont {F.}~\bibnamefont {Nori}},\ }\bibfield  {title}
  {\bibinfo {title} {Enhanced tripartite interactions in spin-magnon-mechanical
  hybrid systems},\ }\href {https://doi.org/10.1103/PhysRevLett.130.073602}
  {\bibfield  {journal} {\bibinfo  {journal} {Phys. Rev. Lett.}\ }\textbf
  {\bibinfo {volume} {130}},\ \bibinfo {pages} {073602} (\bibinfo {year}
  {2023})}\BibitemShut {NoStop}%
\bibitem [{\citenamefont {Moslehi}\ \emph {et~al.}(2023)\citenamefont
  {Moslehi}, \citenamefont {Baghshahi}, \citenamefont {Faghihi},\ and\
  \citenamefont {Mirafzali}}]{Moslehi_2023}%
  \BibitemOpen
  \bibfield  {author} {\bibinfo {author} {\bibfnamefont {M.}~\bibnamefont
  {Moslehi}}, \bibinfo {author} {\bibfnamefont {H.~R.}\ \bibnamefont
  {Baghshahi}}, \bibinfo {author} {\bibfnamefont {M.~J.}\ \bibnamefont
  {Faghihi}},\ and\ \bibinfo {author} {\bibfnamefont {S.~Y.}\ \bibnamefont
  {Mirafzali}},\ }\bibfield  {title} {\bibinfo {title} {Nonclassicality of
  dissipative cavity optomagnonics in the presence of kerr nonlinearities},\
  }\href {https://doi.org/10.1088/1402-4896/acb245} {\bibfield  {journal}
  {\bibinfo  {journal} {Physica Scripta}\ }\textbf {\bibinfo {volume} {98}},\
  \bibinfo {pages} {025103} (\bibinfo {year} {2023})}\BibitemShut {NoStop}%
\bibitem [{\citenamefont {Zhang}\ \emph
  {et~al.}(2019{\natexlab{a}})\citenamefont {Zhang}, \citenamefont {Wang},\
  and\ \citenamefont {You}}]{10.1007/s11433-018-9344-8}%
  \BibitemOpen
  \bibfield  {author} {\bibinfo {author} {\bibfnamefont {G.}~\bibnamefont
  {Zhang}}, \bibinfo {author} {\bibfnamefont {Y.}~\bibnamefont {Wang}},\ and\
  \bibinfo {author} {\bibfnamefont {J.}~\bibnamefont {You}},\ }\bibfield
  {title} {\bibinfo {title} {Theory of the magnon {K}err effect in cavity
  magnonics},\ }\href {https://doi.org/10.1007/s11433-018-9344-8} {\bibfield
  {journal} {\bibinfo  {journal} {Science China Physics, Mechanics Astronomy}\
  }\textbf {\bibinfo {volume} {62}},\ \bibinfo {pages} {987511} (\bibinfo
  {year} {2019}{\natexlab{a}})}\BibitemShut {NoStop}%
\bibitem [{\citenamefont {Wang}\ \emph {et~al.}(2018)\citenamefont {Wang},
  \citenamefont {Zhang}, \citenamefont {Zhang}, \citenamefont {Li},
  \citenamefont {Hu},\ and\ \citenamefont {You}}]{PhysRevLett.120.057202}%
  \BibitemOpen
  \bibfield  {author} {\bibinfo {author} {\bibfnamefont {Y.-P.}\ \bibnamefont
  {Wang}}, \bibinfo {author} {\bibfnamefont {G.-Q.}\ \bibnamefont {Zhang}},
  \bibinfo {author} {\bibfnamefont {D.}~\bibnamefont {Zhang}}, \bibinfo
  {author} {\bibfnamefont {T.-F.}\ \bibnamefont {Li}}, \bibinfo {author}
  {\bibfnamefont {C.-M.}\ \bibnamefont {Hu}},\ and\ \bibinfo {author}
  {\bibfnamefont {J.~Q.}\ \bibnamefont {You}},\ }\bibfield  {title} {\bibinfo
  {title} {Bistability of cavity magnon polaritons},\ }\href
  {https://doi.org/10.1103/PhysRevLett.120.057202} {\bibfield  {journal}
  {\bibinfo  {journal} {Phys. Rev. Lett.}\ }\textbf {\bibinfo {volume} {120}},\
  \bibinfo {pages} {057202} (\bibinfo {year} {2018})}\BibitemShut {NoStop}%
\bibitem [{\citenamefont {Yang}\ \emph {et~al.}(2021)\citenamefont {Yang},
  \citenamefont {Jin}, \citenamefont {Jin}, \citenamefont {Liu}, \citenamefont
  {Liu},\ and\ \citenamefont {Yang}}]{PhysRevResearch.3.023126}%
  \BibitemOpen
  \bibfield  {author} {\bibinfo {author} {\bibfnamefont {Z.-B.}\ \bibnamefont
  {Yang}}, \bibinfo {author} {\bibfnamefont {H.}~\bibnamefont {Jin}}, \bibinfo
  {author} {\bibfnamefont {J.-W.}\ \bibnamefont {Jin}}, \bibinfo {author}
  {\bibfnamefont {J.-Y.}\ \bibnamefont {Liu}}, \bibinfo {author} {\bibfnamefont
  {H.-Y.}\ \bibnamefont {Liu}},\ and\ \bibinfo {author} {\bibfnamefont {R.-C.}\
  \bibnamefont {Yang}},\ }\bibfield  {title} {\bibinfo {title} {Bistability of
  squeezing and entanglement in cavity magnonics},\ }\href
  {https://doi.org/10.1103/PhysRevResearch.3.023126} {\bibfield  {journal}
  {\bibinfo  {journal} {Phys. Rev. Research}\ }\textbf {\bibinfo {volume}
  {3}},\ \bibinfo {pages} {023126} (\bibinfo {year} {2021})}\BibitemShut
  {NoStop}%
\bibitem [{\citenamefont {Shen}\ \emph {et~al.}(2022)\citenamefont {Shen},
  \citenamefont {Li}, \citenamefont {Fan}, \citenamefont {Wang},\ and\
  \citenamefont {You}}]{PhysRevLett.129.123601}%
  \BibitemOpen
  \bibfield  {author} {\bibinfo {author} {\bibfnamefont {R.-C.}\ \bibnamefont
  {Shen}}, \bibinfo {author} {\bibfnamefont {J.}~\bibnamefont {Li}}, \bibinfo
  {author} {\bibfnamefont {Z.-Y.}\ \bibnamefont {Fan}}, \bibinfo {author}
  {\bibfnamefont {Y.-P.}\ \bibnamefont {Wang}},\ and\ \bibinfo {author}
  {\bibfnamefont {J.~Q.}\ \bibnamefont {You}},\ }\bibfield  {title} {\bibinfo
  {title} {Mechanical bistability in {K}err-modified cavity magnomechanics},\
  }\href {https://doi.org/10.1103/PhysRevLett.129.123601} {\bibfield  {journal}
  {\bibinfo  {journal} {Phys. Rev. Lett.}\ }\textbf {\bibinfo {volume} {129}},\
  \bibinfo {pages} {123601} (\bibinfo {year} {2022})}\BibitemShut {NoStop}%
\bibitem [{\citenamefont {Shen}\ \emph {et~al.}(2021)\citenamefont {Shen},
  \citenamefont {Wang}, \citenamefont {Li}, \citenamefont {Zhu}, \citenamefont
  {Agarwal},\ and\ \citenamefont {You}}]{PhysRevLett.127.183202}%
  \BibitemOpen
  \bibfield  {author} {\bibinfo {author} {\bibfnamefont {R.-C.}\ \bibnamefont
  {Shen}}, \bibinfo {author} {\bibfnamefont {Y.-P.}\ \bibnamefont {Wang}},
  \bibinfo {author} {\bibfnamefont {J.}~\bibnamefont {Li}}, \bibinfo {author}
  {\bibfnamefont {S.-Y.}\ \bibnamefont {Zhu}}, \bibinfo {author} {\bibfnamefont
  {G.~S.}\ \bibnamefont {Agarwal}},\ and\ \bibinfo {author} {\bibfnamefont
  {J.~Q.}\ \bibnamefont {You}},\ }\bibfield  {title} {\bibinfo {title}
  {Long-time memory and ternary logic gate using a multistable cavity magnonic
  system},\ }\href {https://doi.org/10.1103/PhysRevLett.127.183202} {\bibfield
  {journal} {\bibinfo  {journal} {Phys. Rev. Lett.}\ }\textbf {\bibinfo
  {volume} {127}},\ \bibinfo {pages} {183202} (\bibinfo {year}
  {2021})}\BibitemShut {NoStop}%
\bibitem [{\citenamefont {Bi}\ \emph {et~al.}(2021)\citenamefont {Bi},
  \citenamefont {Yan}, \citenamefont {Zhang},\ and\ \citenamefont
  {Xiao}}]{PhysRevB.103.104411}%
  \BibitemOpen
  \bibfield  {author} {\bibinfo {author} {\bibfnamefont {M.~X.}\ \bibnamefont
  {Bi}}, \bibinfo {author} {\bibfnamefont {X.~H.}\ \bibnamefont {Yan}},
  \bibinfo {author} {\bibfnamefont {Y.}~\bibnamefont {Zhang}},\ and\ \bibinfo
  {author} {\bibfnamefont {Y.}~\bibnamefont {Xiao}},\ }\bibfield  {title}
  {\bibinfo {title} {Tristability of cavity magnon polaritons},\ }\href
  {https://doi.org/10.1103/PhysRevB.103.104411} {\bibfield  {journal} {\bibinfo
   {journal} {Phys. Rev. B}\ }\textbf {\bibinfo {volume} {103}},\ \bibinfo
  {pages} {104411} (\bibinfo {year} {2021})}\BibitemShut {NoStop}%
\bibitem [{\citenamefont {Kong}\ \emph {et~al.}(2019)\citenamefont {Kong},
  \citenamefont {Xiong},\ and\ \citenamefont {Wu}}]{PhysRevApplied.12.034001}%
  \BibitemOpen
  \bibfield  {author} {\bibinfo {author} {\bibfnamefont {C.}~\bibnamefont
  {Kong}}, \bibinfo {author} {\bibfnamefont {H.}~\bibnamefont {Xiong}},\ and\
  \bibinfo {author} {\bibfnamefont {Y.}~\bibnamefont {Wu}},\ }\bibfield
  {title} {\bibinfo {title} {Magnon-induced nonreciprocity based on the magnon
  {K}err effect},\ }\href {https://doi.org/10.1103/PhysRevApplied.12.034001}
  {\bibfield  {journal} {\bibinfo  {journal} {Phys. Rev. Appl.}\ }\textbf
  {\bibinfo {volume} {12}},\ \bibinfo {pages} {034001} (\bibinfo {year}
  {2019})}\BibitemShut {NoStop}%
\bibitem [{\citenamefont {Liu}\ \emph {et~al.}(2018)\citenamefont {Liu},
  \citenamefont {Wang}, \citenamefont {Xiong},\ and\ \citenamefont
  {Wu}}]{10.1364/OL.43.003698}%
  \BibitemOpen
  \bibfield  {author} {\bibinfo {author} {\bibfnamefont {Z.-X.}\ \bibnamefont
  {Liu}}, \bibinfo {author} {\bibfnamefont {B.}~\bibnamefont {Wang}}, \bibinfo
  {author} {\bibfnamefont {H.}~\bibnamefont {Xiong}},\ and\ \bibinfo {author}
  {\bibfnamefont {Y.}~\bibnamefont {Wu}},\ }\bibfield  {title} {\bibinfo
  {title} {Magnon-induced high-order sideband generation},\ }\href
  {https://doi.org/10.1364/OL.43.003698} {\bibfield  {journal} {\bibinfo
  {journal} {Opt. Lett.}\ }\textbf {\bibinfo {volume} {43}},\ \bibinfo {pages}
  {3698} (\bibinfo {year} {2018})}\BibitemShut {NoStop}%
\bibitem [{\citenamefont {Wang}\ \emph {et~al.}(2021)\citenamefont {Wang},
  \citenamefont {Kong}, \citenamefont {Sun}, \citenamefont {Zhang},
  \citenamefont {Wu},\ and\ \citenamefont {Zheng}}]{PhysRevA.104.033708}%
  \BibitemOpen
  \bibfield  {author} {\bibinfo {author} {\bibfnamefont {M.}~\bibnamefont
  {Wang}}, \bibinfo {author} {\bibfnamefont {C.}~\bibnamefont {Kong}}, \bibinfo
  {author} {\bibfnamefont {Z.-Y.}\ \bibnamefont {Sun}}, \bibinfo {author}
  {\bibfnamefont {D.}~\bibnamefont {Zhang}}, \bibinfo {author} {\bibfnamefont
  {Y.-Y.}\ \bibnamefont {Wu}},\ and\ \bibinfo {author} {\bibfnamefont {L.-L.}\
  \bibnamefont {Zheng}},\ }\bibfield  {title} {\bibinfo {title} {Nonreciprocal
  high-order sidebands induced by magnon {K}err nonlinearity},\ }\href
  {https://doi.org/10.1103/PhysRevA.104.033708} {\bibfield  {journal} {\bibinfo
   {journal} {Phys. Rev. A}\ }\textbf {\bibinfo {volume} {104}},\ \bibinfo
  {pages} {033708} (\bibinfo {year} {2021})}\BibitemShut {NoStop}%
\bibitem [{\citenamefont {Zhao}\ \emph {et~al.}(2022)\citenamefont {Zhao},
  \citenamefont {Yang}, \citenamefont {Peng}, \citenamefont {Yang},
  \citenamefont {Li},\ and\ \citenamefont {Zhou}}]{PhysRevApplied.18.044074}%
  \BibitemOpen
  \bibfield  {author} {\bibinfo {author} {\bibfnamefont {C.}~\bibnamefont
  {Zhao}}, \bibinfo {author} {\bibfnamefont {Z.}~\bibnamefont {Yang}}, \bibinfo
  {author} {\bibfnamefont {R.}~\bibnamefont {Peng}}, \bibinfo {author}
  {\bibfnamefont {J.}~\bibnamefont {Yang}}, \bibinfo {author} {\bibfnamefont
  {C.}~\bibnamefont {Li}},\ and\ \bibinfo {author} {\bibfnamefont
  {L.}~\bibnamefont {Zhou}},\ }\bibfield  {title} {\bibinfo {title}
  {Dissipative-coupling-induced transparency and high-order sidebands with
  {K}err nonlinearity in a cavity-magnonics system},\ }\href
  {https://doi.org/10.1103/PhysRevApplied.18.044074} {\bibfield  {journal}
  {\bibinfo  {journal} {Phys. Rev. Appl.}\ }\textbf {\bibinfo {volume} {18}},\
  \bibinfo {pages} {044074} (\bibinfo {year} {2022})}\BibitemShut {NoStop}%
\bibitem [{\citenamefont {Zhang}\ \emph
  {et~al.}(2019{\natexlab{b}})\citenamefont {Zhang}, \citenamefont {Scully},\
  and\ \citenamefont {Agarwal}}]{PhysRevResearch.1.023021}%
  \BibitemOpen
  \bibfield  {author} {\bibinfo {author} {\bibfnamefont {Z.}~\bibnamefont
  {Zhang}}, \bibinfo {author} {\bibfnamefont {M.~O.}\ \bibnamefont {Scully}},\
  and\ \bibinfo {author} {\bibfnamefont {G.~S.}\ \bibnamefont {Agarwal}},\
  }\bibfield  {title} {\bibinfo {title} {Quantum entanglement between two
  magnon modes via {K}err nonlinearity driven far from equilibrium},\ }\href
  {https://doi.org/10.1103/PhysRevResearch.1.023021} {\bibfield  {journal}
  {\bibinfo  {journal} {Phys. Rev. Research}\ }\textbf {\bibinfo {volume}
  {1}},\ \bibinfo {pages} {023021} (\bibinfo {year}
  {2019}{\natexlab{b}})}\BibitemShut {NoStop}%
\bibitem [{\citenamefont {Qin}\ \emph {et~al.}(2022)\citenamefont {Qin},
  \citenamefont {Li}, \citenamefont {Li},\ and\ \citenamefont
  {Song}}]{PhysRevB.106.054419}%
  \BibitemOpen
  \bibfield  {author} {\bibinfo {author} {\bibfnamefont {Y.}~\bibnamefont
  {Qin}}, \bibinfo {author} {\bibfnamefont {S.-C.}\ \bibnamefont {Li}},
  \bibinfo {author} {\bibfnamefont {K.}~\bibnamefont {Li}},\ and\ \bibinfo
  {author} {\bibfnamefont {J.-J.}\ \bibnamefont {Song}},\ }\bibfield  {title}
  {\bibinfo {title} {Controllable quantum phase transition in a double-cavity
  magnonic system},\ }\href {https://doi.org/10.1103/PhysRevB.106.054419}
  {\bibfield  {journal} {\bibinfo  {journal} {Phys. Rev. B}\ }\textbf {\bibinfo
  {volume} {106}},\ \bibinfo {pages} {054419} (\bibinfo {year}
  {2022})}\BibitemShut {NoStop}%
\bibitem [{\citenamefont {Xiong}\ \emph {et~al.}(2022)\citenamefont {Xiong},
  \citenamefont {Tian}, \citenamefont {Zhang},\ and\ \citenamefont
  {You}}]{PhysRevB.105.245310}%
  \BibitemOpen
  \bibfield  {author} {\bibinfo {author} {\bibfnamefont {W.}~\bibnamefont
  {Xiong}}, \bibinfo {author} {\bibfnamefont {M.}~\bibnamefont {Tian}},
  \bibinfo {author} {\bibfnamefont {G.-Q.}\ \bibnamefont {Zhang}},\ and\
  \bibinfo {author} {\bibfnamefont {J.~Q.}\ \bibnamefont {You}},\ }\bibfield
  {title} {\bibinfo {title} {Strong long-range spin-spin coupling via a {K}err
  magnon interface},\ }\href {https://doi.org/10.1103/PhysRevB.105.245310}
  {\bibfield  {journal} {\bibinfo  {journal} {Phys. Rev. B}\ }\textbf {\bibinfo
  {volume} {105}},\ \bibinfo {pages} {245310} (\bibinfo {year}
  {2022})}\BibitemShut {NoStop}%
\bibitem [{\citenamefont {Li}\ \emph {et~al.}(2018)\citenamefont {Li},
  \citenamefont {Zhu},\ and\ \citenamefont {Agarwal}}]{PhysRevLett.121.203601}%
  \BibitemOpen
  \bibfield  {author} {\bibinfo {author} {\bibfnamefont {J.}~\bibnamefont
  {Li}}, \bibinfo {author} {\bibfnamefont {S.-Y.}\ \bibnamefont {Zhu}},\ and\
  \bibinfo {author} {\bibfnamefont {G.~S.}\ \bibnamefont {Agarwal}},\
  }\bibfield  {title} {\bibinfo {title} {Magnon-photon-phonon entanglement in
  cavity magnomechanics},\ }\href
  {https://doi.org/10.1103/PhysRevLett.121.203601} {\bibfield  {journal}
  {\bibinfo  {journal} {Phys. Rev. Lett.}\ }\textbf {\bibinfo {volume} {121}},\
  \bibinfo {pages} {203601} (\bibinfo {year} {2018})}\BibitemShut {NoStop}%
\bibitem [{\citenamefont {An}\ \emph {et~al.}(2020)\citenamefont {An},
  \citenamefont {Litvinenko}, \citenamefont {Kohno}, \citenamefont {Fuad},
  \citenamefont {Naletov}, \citenamefont {Vila}, \citenamefont {Ebels},
  \citenamefont {de~Loubens}, \citenamefont {Hurdequint}, \citenamefont
  {Beaulieu}, \citenamefont {Ben~Youssef}, \citenamefont {Vukadinovic},
  \citenamefont {Bauer}, \citenamefont {Slavin}, \citenamefont {Tiberkevich},\
  and\ \citenamefont {Klein}}]{PhysRevB.101.060407}%
  \BibitemOpen
  \bibfield  {author} {\bibinfo {author} {\bibfnamefont {K.}~\bibnamefont
  {An}}, \bibinfo {author} {\bibfnamefont {A.~N.}\ \bibnamefont {Litvinenko}},
  \bibinfo {author} {\bibfnamefont {R.}~\bibnamefont {Kohno}}, \bibinfo
  {author} {\bibfnamefont {A.~A.}\ \bibnamefont {Fuad}}, \bibinfo {author}
  {\bibfnamefont {V.~V.}\ \bibnamefont {Naletov}}, \bibinfo {author}
  {\bibfnamefont {L.}~\bibnamefont {Vila}}, \bibinfo {author} {\bibfnamefont
  {U.}~\bibnamefont {Ebels}}, \bibinfo {author} {\bibfnamefont
  {G.}~\bibnamefont {de~Loubens}}, \bibinfo {author} {\bibfnamefont
  {H.}~\bibnamefont {Hurdequint}}, \bibinfo {author} {\bibfnamefont
  {N.}~\bibnamefont {Beaulieu}}, \bibinfo {author} {\bibfnamefont
  {J.}~\bibnamefont {Ben~Youssef}}, \bibinfo {author} {\bibfnamefont
  {N.}~\bibnamefont {Vukadinovic}}, \bibinfo {author} {\bibfnamefont
  {G.~E.~W.}\ \bibnamefont {Bauer}}, \bibinfo {author} {\bibfnamefont {A.~N.}\
  \bibnamefont {Slavin}}, \bibinfo {author} {\bibfnamefont {V.~S.}\
  \bibnamefont {Tiberkevich}},\ and\ \bibinfo {author} {\bibfnamefont
  {O.}~\bibnamefont {Klein}},\ }\bibfield  {title} {\bibinfo {title} {Coherent
  long-range transfer of angular momentum between magnon {K}ittel modes by
  phonons},\ }\href {https://doi.org/10.1103/PhysRevB.101.060407} {\bibfield
  {journal} {\bibinfo  {journal} {Phys. Rev. B}\ }\textbf {\bibinfo {volume}
  {101}},\ \bibinfo {pages} {060407(R)} (\bibinfo {year} {2020})}\BibitemShut
  {NoStop}%
\bibitem [{\citenamefont {Qi}\ and\ \citenamefont
  {Jing}(2021)}]{PhysRevA.104.032606}%
  \BibitemOpen
  \bibfield  {author} {\bibinfo {author} {\bibfnamefont {S.-f.}\ \bibnamefont
  {Qi}}\ and\ \bibinfo {author} {\bibfnamefont {J.}~\bibnamefont {Jing}},\
  }\bibfield  {title} {\bibinfo {title} {Magnon-mediated quantum battery under
  systematic errors},\ }\href {https://doi.org/10.1103/PhysRevA.104.032606}
  {\bibfield  {journal} {\bibinfo  {journal} {Phys. Rev. A}\ }\textbf {\bibinfo
  {volume} {104}},\ \bibinfo {pages} {032606} (\bibinfo {year}
  {2021})}\BibitemShut {NoStop}%
\bibitem [{\citenamefont {Ren}\ \emph {et~al.}(2022)\citenamefont {Ren},
  \citenamefont {Xie}, \citenamefont {Li}, \citenamefont {Ma},\ and\
  \citenamefont {Li}}]{PhysRevB.105.094422}%
  \BibitemOpen
  \bibfield  {author} {\bibinfo {author} {\bibfnamefont {Y.-l.}\ \bibnamefont
  {Ren}}, \bibinfo {author} {\bibfnamefont {J.-k.}\ \bibnamefont {Xie}},
  \bibinfo {author} {\bibfnamefont {X.-k.}\ \bibnamefont {Li}}, \bibinfo
  {author} {\bibfnamefont {S.-l.}\ \bibnamefont {Ma}},\ and\ \bibinfo {author}
  {\bibfnamefont {F.-l.}\ \bibnamefont {Li}},\ }\bibfield  {title} {\bibinfo
  {title} {Long-range generation of a magnon-magnon entangled state},\ }\href
  {https://doi.org/10.1103/PhysRevB.105.094422} {\bibfield  {journal} {\bibinfo
   {journal} {Phys. Rev. B}\ }\textbf {\bibinfo {volume} {105}},\ \bibinfo
  {pages} {094422} (\bibinfo {year} {2022})}\BibitemShut {NoStop}%
\bibitem [{\citenamefont {Morris}\ \emph {et~al.}(2017)\citenamefont {Morris},
  \citenamefont {van Loo}, \citenamefont {Kosen},\ and\ \citenamefont
  {Karenowska}}]{10.1038/s41598-017-11835-4}%
  \BibitemOpen
  \bibfield  {author} {\bibinfo {author} {\bibfnamefont {R.~G.~E.}\
  \bibnamefont {Morris}}, \bibinfo {author} {\bibfnamefont {A.~F.}\
  \bibnamefont {van Loo}}, \bibinfo {author} {\bibfnamefont {S.}~\bibnamefont
  {Kosen}},\ and\ \bibinfo {author} {\bibfnamefont {A.~D.}\ \bibnamefont
  {Karenowska}},\ }\bibfield  {title} {\bibinfo {title} {Strong coupling of
  magnons in a {YIG} sphere to photons in a planar superconducting resonator in
  the quantum limit},\ }\href {https://doi.org/10.1038/s41598-017-11835-4}
  {\bibfield  {journal} {\bibinfo  {journal} {Scientific Reports}\ }\textbf
  {\bibinfo {volume} {7}},\ \bibinfo {pages} {11511} (\bibinfo {year}
  {2017})}\BibitemShut {NoStop}%
\bibitem [{\citenamefont {Rao}\ \emph {et~al.}(2023)\citenamefont {Rao},
  \citenamefont {Yao}, \citenamefont {Wang}, \citenamefont {Zhang},
  \citenamefont {Yu},\ and\ \citenamefont {Lu}}]{PhysRevLett.130.046705}%
  \BibitemOpen
  \bibfield  {author} {\bibinfo {author} {\bibfnamefont {J.~W.}\ \bibnamefont
  {Rao}}, \bibinfo {author} {\bibfnamefont {B.}~\bibnamefont {Yao}}, \bibinfo
  {author} {\bibfnamefont {C.~Y.}\ \bibnamefont {Wang}}, \bibinfo {author}
  {\bibfnamefont {C.}~\bibnamefont {Zhang}}, \bibinfo {author} {\bibfnamefont
  {T.}~\bibnamefont {Yu}},\ and\ \bibinfo {author} {\bibfnamefont
  {W.}~\bibnamefont {Lu}},\ }\bibfield  {title} {\bibinfo {title} {Unveiling a
  pump-induced magnon mode via its strong interaction with walker modes},\
  }\href {https://doi.org/10.1103/PhysRevLett.130.046705} {\bibfield  {journal}
  {\bibinfo  {journal} {Phys. Rev. Lett.}\ }\textbf {\bibinfo {volume} {130}},\
  \bibinfo {pages} {046705} (\bibinfo {year} {2023})}\BibitemShut {NoStop}%
\bibitem [{\citenamefont {Bittencourt}\ \emph {et~al.}(2023)\citenamefont
  {Bittencourt}, \citenamefont {Potts}, \citenamefont {Huang}, \citenamefont
  {Davis},\ and\ \citenamefont {Viola~Kusminskiy}}]{PhysRevB.107.144411}%
  \BibitemOpen
  \bibfield  {author} {\bibinfo {author} {\bibfnamefont {V.~A. S.~V.}\
  \bibnamefont {Bittencourt}}, \bibinfo {author} {\bibfnamefont {C.~A.}\
  \bibnamefont {Potts}}, \bibinfo {author} {\bibfnamefont {Y.}~\bibnamefont
  {Huang}}, \bibinfo {author} {\bibfnamefont {J.~P.}\ \bibnamefont {Davis}},\
  and\ \bibinfo {author} {\bibfnamefont {S.}~\bibnamefont {Viola~Kusminskiy}},\
  }\bibfield  {title} {\bibinfo {title} {Magnomechanical backaction corrections
  due to coupling to higher-order walker modes and {K}err nonlinearities},\
  }\href {https://doi.org/10.1103/PhysRevB.107.144411} {\bibfield  {journal}
  {\bibinfo  {journal} {Phys. Rev. B}\ }\textbf {\bibinfo {volume} {107}},\
  \bibinfo {pages} {144411} (\bibinfo {year} {2023})}\BibitemShut {NoStop}%
\bibitem [{\citenamefont {Walker}(2004)}]{10.1063/1.1723117}%
  \BibitemOpen
  \bibfield  {author} {\bibinfo {author} {\bibfnamefont {L.~R.}\ \bibnamefont
  {Walker}},\ }\bibfield  {title} {\bibinfo {title} {Resonant modes of
  ferromagnetic spheroids},\ }\href {https://doi.org/10.1063/1.1723117 %J
  Journal of Applied Physics} {\bibfield  {journal} {\bibinfo  {journal}
  {Journal of Applied Physics}\ }\textbf {\bibinfo {volume} {29}},\ \bibinfo
  {pages} {318} (\bibinfo {year} {2004})}\BibitemShut {NoStop}%
\bibitem [{\citenamefont {Elyasi}\ \emph {et~al.}(2020)\citenamefont {Elyasi},
  \citenamefont {Blanter},\ and\ \citenamefont {Bauer}}]{PhysRevB.101.054402}%
  \BibitemOpen
  \bibfield  {author} {\bibinfo {author} {\bibfnamefont {M.}~\bibnamefont
  {Elyasi}}, \bibinfo {author} {\bibfnamefont {Y.~M.}\ \bibnamefont
  {Blanter}},\ and\ \bibinfo {author} {\bibfnamefont {G.~E.~W.}\ \bibnamefont
  {Bauer}},\ }\bibfield  {title} {\bibinfo {title} {Resources of nonlinear
  cavity magnonics for quantum information},\ }\href
  {https://doi.org/10.1103/PhysRevB.101.054402} {\bibfield  {journal} {\bibinfo
   {journal} {Phys. Rev. B}\ }\textbf {\bibinfo {volume} {101}},\ \bibinfo
  {pages} {054402} (\bibinfo {year} {2020})}\BibitemShut {NoStop}%
\bibitem [{SM()}]{SM}%
  \BibitemOpen
  \href@noop {} {}\bibinfo {note} {See the Supplemental Material for a detailed
  derivation of the linearized Hamiltonian, the magnetostatic Green’s tensor,
  the spectral density, the energy spectrum, and the exact master equation of
  the spin system and its steady-state solution.}\BibitemShut {Stop}%
\bibitem [{\citenamefont {Crescini}\ \emph {et~al.}(2021)\citenamefont
  {Crescini}, \citenamefont {Braggio}, \citenamefont {Carugno}, \citenamefont
  {Ortolan},\ and\ \citenamefont {Ruoso}}]{PhysRevB.104.064426}%
  \BibitemOpen
  \bibfield  {author} {\bibinfo {author} {\bibfnamefont {N.}~\bibnamefont
  {Crescini}}, \bibinfo {author} {\bibfnamefont {C.}~\bibnamefont {Braggio}},
  \bibinfo {author} {\bibfnamefont {G.}~\bibnamefont {Carugno}}, \bibinfo
  {author} {\bibfnamefont {A.}~\bibnamefont {Ortolan}},\ and\ \bibinfo {author}
  {\bibfnamefont {G.}~\bibnamefont {Ruoso}},\ }\bibfield  {title} {\bibinfo
  {title} {Coherent coupling between multiple ferrimagnetic spheres and a
  microwave cavity at millikelvin temperatures},\ }\href
  {https://doi.org/10.1103/PhysRevB.104.064426} {\bibfield  {journal} {\bibinfo
   {journal} {Phys. Rev. B}\ }\textbf {\bibinfo {volume} {104}},\ \bibinfo
  {pages} {064426} (\bibinfo {year} {2021})}\BibitemShut {NoStop}%
\bibitem [{\citenamefont {Mills}\ and\ \citenamefont
  {Burstein}(1974)}]{Mills1974}%
  \BibitemOpen
  \bibfield  {author} {\bibinfo {author} {\bibfnamefont {D.~L.}\ \bibnamefont
  {Mills}}\ and\ \bibinfo {author} {\bibfnamefont {E.}~\bibnamefont
  {Burstein}},\ }\bibfield  {title} {\bibinfo {title} {Polaritons: the
  electromagnetic modes of media},\ }\href
  {https://doi.org/10.1088/0034-4885/37/7/001} {\bibfield  {journal} {\bibinfo
  {journal} {Reports on Progress in Physics}\ }\textbf {\bibinfo {volume}
  {37}},\ \bibinfo {pages} {817} (\bibinfo {year} {1974})}\BibitemShut
  {NoStop}%
\bibitem [{\citenamefont {Walker}(1957)}]{PhysRev.105.390}%
  \BibitemOpen
  \bibfield  {author} {\bibinfo {author} {\bibfnamefont {L.~R.}\ \bibnamefont
  {Walker}},\ }\bibfield  {title} {\bibinfo {title} {Magnetostatic modes in
  ferromagnetic resonance},\ }\href {https://doi.org/10.1103/PhysRev.105.390}
  {\bibfield  {journal} {\bibinfo  {journal} {Phys. Rev.}\ }\textbf {\bibinfo
  {volume} {105}},\ \bibinfo {pages} {390} (\bibinfo {year}
  {1957})}\BibitemShut {NoStop}%
\bibitem [{\citenamefont {Fletcher}\ and\ \citenamefont
  {Bell}(1959)}]{10.1063/1.1735216}%
  \BibitemOpen
  \bibfield  {author} {\bibinfo {author} {\bibfnamefont {P.~C.}\ \bibnamefont
  {Fletcher}}\ and\ \bibinfo {author} {\bibfnamefont {R.~O.}\ \bibnamefont
  {Bell}},\ }\bibfield  {title} {\bibinfo {title} {Ferrimagnetic resonance
  modes in spheres},\ }\href {https://doi.org/10.1063/1.1735216} {\bibfield
  {journal} {\bibinfo  {journal} {Journal of Applied Physics}\ }\textbf
  {\bibinfo {volume} {30}},\ \bibinfo {pages} {687} (\bibinfo {year}
  {1959})}\BibitemShut {NoStop}%
\bibitem [{\citenamefont {Zhao}\ \emph {et~al.}(2021)\citenamefont {Zhao},
  \citenamefont {Wang},\ and\ \citenamefont {Qian}}]{PhysRevB.104.134423}%
  \BibitemOpen
  \bibfield  {author} {\bibinfo {author} {\bibfnamefont {G.}~\bibnamefont
  {Zhao}}, \bibinfo {author} {\bibfnamefont {Y.}~\bibnamefont {Wang}},\ and\
  \bibinfo {author} {\bibfnamefont {X.-F.}\ \bibnamefont {Qian}},\ }\bibfield
  {title} {\bibinfo {title} {Driven dissipative quantum dynamics in a cavity
  magnon-polariton system},\ }\href
  {https://doi.org/10.1103/PhysRevB.104.134423} {\bibfield  {journal} {\bibinfo
   {journal} {Phys. Rev. B}\ }\textbf {\bibinfo {volume} {104}},\ \bibinfo
  {pages} {134423} (\bibinfo {year} {2021})}\BibitemShut {NoStop}%
\bibitem [{\citenamefont {Nair}\ \emph {et~al.}(2021)\citenamefont {Nair},
  \citenamefont {Mukhopadhyay},\ and\ \citenamefont
  {Agarwal}}]{PhysRevB.103.224401}%
  \BibitemOpen
  \bibfield  {author} {\bibinfo {author} {\bibfnamefont {J.~M.~P.}\
  \bibnamefont {Nair}}, \bibinfo {author} {\bibfnamefont {D.}~\bibnamefont
  {Mukhopadhyay}},\ and\ \bibinfo {author} {\bibfnamefont {G.~S.}\ \bibnamefont
  {Agarwal}},\ }\bibfield  {title} {\bibinfo {title} {Ultralow threshold
  bistability and generation of long-lived mode in a dissipatively coupled
  nonlinear system: Application to magnonics},\ }\href
  {https://doi.org/10.1103/PhysRevB.103.224401} {\bibfield  {journal} {\bibinfo
   {journal} {Phys. Rev. B}\ }\textbf {\bibinfo {volume} {103}},\ \bibinfo
  {pages} {224401} (\bibinfo {year} {2021})}\BibitemShut {NoStop}%
\bibitem [{\citenamefont {Zhang}\ \emph {et~al.}(2021)\citenamefont {Zhang},
  \citenamefont {Chen}, \citenamefont {Xiong}, \citenamefont {Lam},\ and\
  \citenamefont {You}}]{PhysRevB.104.064423}%
  \BibitemOpen
  \bibfield  {author} {\bibinfo {author} {\bibfnamefont {G.-Q.}\ \bibnamefont
  {Zhang}}, \bibinfo {author} {\bibfnamefont {Z.}~\bibnamefont {Chen}},
  \bibinfo {author} {\bibfnamefont {W.}~\bibnamefont {Xiong}}, \bibinfo
  {author} {\bibfnamefont {C.-H.}\ \bibnamefont {Lam}},\ and\ \bibinfo {author}
  {\bibfnamefont {J.~Q.}\ \bibnamefont {You}},\ }\bibfield  {title} {\bibinfo
  {title} {Parity-symmetry-breaking quantum phase transition via parametric
  drive in a cavity magnonic system},\ }\href
  {https://doi.org/10.1103/PhysRevB.104.064423} {\bibfield  {journal} {\bibinfo
   {journal} {Phys. Rev. B}\ }\textbf {\bibinfo {volume} {104}},\ \bibinfo
  {pages} {064423} (\bibinfo {year} {2021})}\BibitemShut {NoStop}%
\bibitem [{\citenamefont {Delga}\ \emph {et~al.}(2014)\citenamefont {Delga},
  \citenamefont {Feist}, \citenamefont {Bravo-Abad},\ and\ \citenamefont
  {Garcia-Vidal}}]{PhysRevLett.112.253601}%
  \BibitemOpen
  \bibfield  {author} {\bibinfo {author} {\bibfnamefont {A.}~\bibnamefont
  {Delga}}, \bibinfo {author} {\bibfnamefont {J.}~\bibnamefont {Feist}},
  \bibinfo {author} {\bibfnamefont {J.}~\bibnamefont {Bravo-Abad}},\ and\
  \bibinfo {author} {\bibfnamefont {F.~J.}\ \bibnamefont {Garcia-Vidal}},\
  }\bibfield  {title} {\bibinfo {title} {Quantum emitters near a metal
  nanoparticle: Strong coupling and quenching},\ }\href
  {https://doi.org/10.1103/PhysRevLett.112.253601} {\bibfield  {journal}
  {\bibinfo  {journal} {Phys. Rev. Lett.}\ }\textbf {\bibinfo {volume} {112}},\
  \bibinfo {pages} {253601} (\bibinfo {year} {2014})}\BibitemShut {NoStop}%
\bibitem [{\citenamefont {Gonz\'alez-Tudela}\ \emph {et~al.}(2014)\citenamefont
  {Gonz\'alez-Tudela}, \citenamefont {Huidobro}, \citenamefont
  {Mart\'{\i}n-Moreno}, \citenamefont {Tejedor},\ and\ \citenamefont
  {Garc\'{\i}a-Vidal}}]{PhysRevB.89.041402}%
  \BibitemOpen
  \bibfield  {author} {\bibinfo {author} {\bibfnamefont {A.}~\bibnamefont
  {Gonz\'alez-Tudela}}, \bibinfo {author} {\bibfnamefont {P.~A.}\ \bibnamefont
  {Huidobro}}, \bibinfo {author} {\bibfnamefont {L.}~\bibnamefont
  {Mart\'{\i}n-Moreno}}, \bibinfo {author} {\bibfnamefont {C.}~\bibnamefont
  {Tejedor}},\ and\ \bibinfo {author} {\bibfnamefont {F.~J.}\ \bibnamefont
  {Garc\'{\i}a-Vidal}},\ }\bibfield  {title} {\bibinfo {title} {Reversible
  dynamics of single quantum emitters near metal-dielectric interfaces},\
  }\href {https://doi.org/10.1103/PhysRevB.89.041402} {\bibfield  {journal}
  {\bibinfo  {journal} {Phys. Rev. B}\ }\textbf {\bibinfo {volume} {89}},\
  \bibinfo {pages} {041402(R)} (\bibinfo {year} {2014})}\BibitemShut {NoStop}%
\bibitem [{\citenamefont {Yang}\ and\ \citenamefont
  {An}(2017)}]{PhysRevB.95.161408}%
  \BibitemOpen
  \bibfield  {author} {\bibinfo {author} {\bibfnamefont {C.-J.}\ \bibnamefont
  {Yang}}\ and\ \bibinfo {author} {\bibfnamefont {J.-H.}\ \bibnamefont {An}},\
  }\bibfield  {title} {\bibinfo {title} {Suppressed dissipation of a quantum
  emitter coupled to surface plasmon polaritons},\ }\href
  {https://doi.org/10.1103/PhysRevB.95.161408} {\bibfield  {journal} {\bibinfo
  {journal} {Phys. Rev. B}\ }\textbf {\bibinfo {volume} {95}},\ \bibinfo
  {pages} {161408(R)} (\bibinfo {year} {2017})}\BibitemShut {NoStop}%
\bibitem [{\citenamefont {Tamascelli}\ \emph {et~al.}(2018)\citenamefont
  {Tamascelli}, \citenamefont {Smirne}, \citenamefont {Huelga},\ and\
  \citenamefont {Plenio}}]{PhysRevLett.120.030402}%
  \BibitemOpen
  \bibfield  {author} {\bibinfo {author} {\bibfnamefont {D.}~\bibnamefont
  {Tamascelli}}, \bibinfo {author} {\bibfnamefont {A.}~\bibnamefont {Smirne}},
  \bibinfo {author} {\bibfnamefont {S.~F.}\ \bibnamefont {Huelga}},\ and\
  \bibinfo {author} {\bibfnamefont {M.~B.}\ \bibnamefont {Plenio}},\ }\bibfield
   {title} {\bibinfo {title} {Nonperturbative treatment of non-{M}arkovian
  dynamics of open quantum systems},\ }\href
  {https://doi.org/10.1103/PhysRevLett.120.030402} {\bibfield  {journal}
  {\bibinfo  {journal} {Phys. Rev. Lett.}\ }\textbf {\bibinfo {volume} {120}},\
  \bibinfo {pages} {030402} (\bibinfo {year} {2018})}\BibitemShut {NoStop}%
\bibitem [{\citenamefont {Medina}\ \emph {et~al.}(2021)\citenamefont {Medina},
  \citenamefont {Garc\'{\i}a-Vidal}, \citenamefont
  {Fern\'andez-Dom\'{\i}nguez},\ and\ \citenamefont
  {Feist}}]{PhysRevLett.126.093601}%
  \BibitemOpen
  \bibfield  {author} {\bibinfo {author} {\bibfnamefont {I.}~\bibnamefont
  {Medina}}, \bibinfo {author} {\bibfnamefont {F.~J.}\ \bibnamefont
  {Garc\'{\i}a-Vidal}}, \bibinfo {author} {\bibfnamefont {A.~I.}\ \bibnamefont
  {Fern\'andez-Dom\'{\i}nguez}},\ and\ \bibinfo {author} {\bibfnamefont
  {J.}~\bibnamefont {Feist}},\ }\bibfield  {title} {\bibinfo {title} {Few-mode
  field quantization of arbitrary electromagnetic spectral densities},\ }\href
  {https://doi.org/10.1103/PhysRevLett.126.093601} {\bibfield  {journal}
  {\bibinfo  {journal} {Phys. Rev. Lett.}\ }\textbf {\bibinfo {volume} {126}},\
  \bibinfo {pages} {093601} (\bibinfo {year} {2021})}\BibitemShut {NoStop}%
\bibitem [{\citenamefont {Ji}\ \emph {et~al.}(2022)\citenamefont {Ji},
  \citenamefont {Bai},\ and\ \citenamefont {An}}]{PhysRevB.106.115427}%
  \BibitemOpen
  \bibfield  {author} {\bibinfo {author} {\bibfnamefont {F.-Z.}\ \bibnamefont
  {Ji}}, \bibinfo {author} {\bibfnamefont {S.-Y.}\ \bibnamefont {Bai}},\ and\
  \bibinfo {author} {\bibfnamefont {J.-H.}\ \bibnamefont {An}},\ }\bibfield
  {title} {\bibinfo {title} {Strong coupling of quantum emitters and the
  exciton polariton in {M}o{S}$_{2}$ nanodisks},\ }\href
  {https://doi.org/10.1103/PhysRevB.106.115427} {\bibfield  {journal} {\bibinfo
   {journal} {Phys. Rev. B}\ }\textbf {\bibinfo {volume} {106}},\ \bibinfo
  {pages} {115427} (\bibinfo {year} {2022})}\BibitemShut {NoStop}%
\bibitem [{\citenamefont {Baranov}\ \emph {et~al.}(2018)\citenamefont
  {Baranov}, \citenamefont {Wersäll}, \citenamefont {Cuadra}, \citenamefont
  {Antosiewicz},\ and\ \citenamefont {Shegai}}]{acsphotonics2018}%
  \BibitemOpen
  \bibfield  {author} {\bibinfo {author} {\bibfnamefont {D.~G.}\ \bibnamefont
  {Baranov}}, \bibinfo {author} {\bibfnamefont {M.}~\bibnamefont {Wersäll}},
  \bibinfo {author} {\bibfnamefont {J.}~\bibnamefont {Cuadra}}, \bibinfo
  {author} {\bibfnamefont {T.~J.}\ \bibnamefont {Antosiewicz}},\ and\ \bibinfo
  {author} {\bibfnamefont {T.}~\bibnamefont {Shegai}},\ }\bibfield  {title}
  {\bibinfo {title} {Novel nanostructures and materials for strong
  light–matter interactions},\ }\href
  {https://doi.org/10.1021/acsphotonics.7b00674} {\bibfield  {journal}
  {\bibinfo  {journal} {ACS Photonics}\ }\textbf {\bibinfo {volume} {5}},\
  \bibinfo {pages} {24} (\bibinfo {year} {2018})}\BibitemShut {NoStop}%
\bibitem [{\citenamefont {Sloan}\ \emph {et~al.}(2019)\citenamefont {Sloan},
  \citenamefont {Rivera}, \citenamefont {Joannopoulos}, \citenamefont
  {Kaminer},\ and\ \citenamefont {Solja\ifmmode \check{c}\else
  \v{c}\fi{}i\ifmmode~\acute{c}\else \'{c}\fi{}}}]{PhysRevB.100.235453}%
  \BibitemOpen
  \bibfield  {author} {\bibinfo {author} {\bibfnamefont {J.}~\bibnamefont
  {Sloan}}, \bibinfo {author} {\bibfnamefont {N.}~\bibnamefont {Rivera}},
  \bibinfo {author} {\bibfnamefont {J.~D.}\ \bibnamefont {Joannopoulos}},
  \bibinfo {author} {\bibfnamefont {I.}~\bibnamefont {Kaminer}},\ and\ \bibinfo
  {author} {\bibfnamefont {M.}~\bibnamefont {Solja\ifmmode \check{c}\else
  \v{c}\fi{}i\ifmmode~\acute{c}\else \'{c}\fi{}}},\ }\bibfield  {title}
  {\bibinfo {title} {Controlling spins with surface magnon polaritons},\ }\href
  {https://doi.org/10.1103/PhysRevB.100.235453} {\bibfield  {journal} {\bibinfo
   {journal} {Phys. Rev. B}\ }\textbf {\bibinfo {volume} {100}},\ \bibinfo
  {pages} {235453} (\bibinfo {year} {2019})}\BibitemShut {NoStop}%
\bibitem [{\citenamefont {Tabuchi}\ \emph {et~al.}(2014)\citenamefont
  {Tabuchi}, \citenamefont {Ishino}, \citenamefont {Ishikawa}, \citenamefont
  {Yamazaki}, \citenamefont {Usami},\ and\ \citenamefont
  {Nakamura}}]{PhysRevLett.113.083603}%
  \BibitemOpen
  \bibfield  {author} {\bibinfo {author} {\bibfnamefont {Y.}~\bibnamefont
  {Tabuchi}}, \bibinfo {author} {\bibfnamefont {S.}~\bibnamefont {Ishino}},
  \bibinfo {author} {\bibfnamefont {T.}~\bibnamefont {Ishikawa}}, \bibinfo
  {author} {\bibfnamefont {R.}~\bibnamefont {Yamazaki}}, \bibinfo {author}
  {\bibfnamefont {K.}~\bibnamefont {Usami}},\ and\ \bibinfo {author}
  {\bibfnamefont {Y.}~\bibnamefont {Nakamura}},\ }\bibfield  {title} {\bibinfo
  {title} {Hybridizing ferromagnetic magnons and microwave photons in the
  quantum limit},\ }\href {https://doi.org/10.1103/PhysRevLett.113.083603}
  {\bibfield  {journal} {\bibinfo  {journal} {Phys. Rev. Lett.}\ }\textbf
  {\bibinfo {volume} {113}},\ \bibinfo {pages} {083603} (\bibinfo {year}
  {2014})}\BibitemShut {NoStop}%
\bibitem [{\citenamefont {Zhang}\ \emph {et~al.}(2017)\citenamefont {Zhang},
  \citenamefont {Luo}, \citenamefont {Wang}, \citenamefont {Li},\ and\
  \citenamefont {You}}]{RN1}%
  \BibitemOpen
  \bibfield  {author} {\bibinfo {author} {\bibfnamefont {D.}~\bibnamefont
  {Zhang}}, \bibinfo {author} {\bibfnamefont {X.-Q.}\ \bibnamefont {Luo}},
  \bibinfo {author} {\bibfnamefont {Y.-P.}\ \bibnamefont {Wang}}, \bibinfo
  {author} {\bibfnamefont {T.-F.}\ \bibnamefont {Li}},\ and\ \bibinfo {author}
  {\bibfnamefont {J.~Q.}\ \bibnamefont {You}},\ }\bibfield  {title} {\bibinfo
  {title} {Observation of the exceptional point in cavity magnon-polaritons},\
  }\href {https://doi.org/10.1038/s41467-017-01634-w} {\bibfield  {journal}
  {\bibinfo  {journal} {Nature Communications}\ }\textbf {\bibinfo {volume}
  {8}},\ \bibinfo {pages} {1368} (\bibinfo {year} {2017})}\BibitemShut
  {NoStop}%
\bibitem [{\citenamefont {Huebl}\ \emph {et~al.}(2013)\citenamefont {Huebl},
  \citenamefont {Zollitsch}, \citenamefont {Lotze}, \citenamefont {Hocke},
  \citenamefont {Greifenstein}, \citenamefont {Marx}, \citenamefont {Gross},\
  and\ \citenamefont {Goennenwein}}]{PhysRevLett.111.127003}%
  \BibitemOpen
  \bibfield  {author} {\bibinfo {author} {\bibfnamefont {H.}~\bibnamefont
  {Huebl}}, \bibinfo {author} {\bibfnamefont {C.~W.}\ \bibnamefont
  {Zollitsch}}, \bibinfo {author} {\bibfnamefont {J.}~\bibnamefont {Lotze}},
  \bibinfo {author} {\bibfnamefont {F.}~\bibnamefont {Hocke}}, \bibinfo
  {author} {\bibfnamefont {M.}~\bibnamefont {Greifenstein}}, \bibinfo {author}
  {\bibfnamefont {A.}~\bibnamefont {Marx}}, \bibinfo {author} {\bibfnamefont
  {R.}~\bibnamefont {Gross}},\ and\ \bibinfo {author} {\bibfnamefont
  {S.~T.~B.}\ \bibnamefont {Goennenwein}},\ }\bibfield  {title} {\bibinfo
  {title} {High cooperativity in coupled microwave resonator ferrimagnetic
  insulator hybrids},\ }\href {https://doi.org/10.1103/PhysRevLett.111.127003}
  {\bibfield  {journal} {\bibinfo  {journal} {Phys. Rev. Lett.}\ }\textbf
  {\bibinfo {volume} {111}},\ \bibinfo {pages} {127003} (\bibinfo {year}
  {2013})}\BibitemShut {NoStop}%
\bibitem [{\citenamefont {Wang}\ \emph {et~al.}(2016)\citenamefont {Wang},
  \citenamefont {Zhang}, \citenamefont {Zhang}, \citenamefont {Luo},
  \citenamefont {Xiong}, \citenamefont {Wang}, \citenamefont {Li},
  \citenamefont {Hu},\ and\ \citenamefont {You}}]{PhysRevB.94.224410}%
  \BibitemOpen
  \bibfield  {author} {\bibinfo {author} {\bibfnamefont {Y.-P.}\ \bibnamefont
  {Wang}}, \bibinfo {author} {\bibfnamefont {G.-Q.}\ \bibnamefont {Zhang}},
  \bibinfo {author} {\bibfnamefont {D.}~\bibnamefont {Zhang}}, \bibinfo
  {author} {\bibfnamefont {X.-Q.}\ \bibnamefont {Luo}}, \bibinfo {author}
  {\bibfnamefont {W.}~\bibnamefont {Xiong}}, \bibinfo {author} {\bibfnamefont
  {S.-P.}\ \bibnamefont {Wang}}, \bibinfo {author} {\bibfnamefont {T.-F.}\
  \bibnamefont {Li}}, \bibinfo {author} {\bibfnamefont {C.-M.}\ \bibnamefont
  {Hu}},\ and\ \bibinfo {author} {\bibfnamefont {J.~Q.}\ \bibnamefont {You}},\
  }\bibfield  {title} {\bibinfo {title} {Magnon {K}err effect in a strongly
  coupled cavity-magnon system},\ }\href
  {https://doi.org/10.1103/PhysRevB.94.224410} {\bibfield  {journal} {\bibinfo
  {journal} {Phys. Rev. B}\ }\textbf {\bibinfo {volume} {94}},\ \bibinfo
  {pages} {224410} (\bibinfo {year} {2016})}\BibitemShut {NoStop}%
\bibitem [{\citenamefont {Li}\ \emph {et~al.}(2022{\natexlab{b}})\citenamefont
  {Li}, \citenamefont {Yefremenko}, \citenamefont {Lisovenko}, \citenamefont
  {Trevillian}, \citenamefont {Polakovic}, \citenamefont {Cecil}, \citenamefont
  {Barry}, \citenamefont {Pearson}, \citenamefont {Divan}, \citenamefont
  {Tyberkevych}, \citenamefont {Chang}, \citenamefont {Welp}, \citenamefont
  {Kwok},\ and\ \citenamefont {Novosad}}]{PhysRevLett.128.047701}%
  \BibitemOpen
  \bibfield  {author} {\bibinfo {author} {\bibfnamefont {Y.}~\bibnamefont
  {Li}}, \bibinfo {author} {\bibfnamefont {V.~G.}\ \bibnamefont {Yefremenko}},
  \bibinfo {author} {\bibfnamefont {M.}~\bibnamefont {Lisovenko}}, \bibinfo
  {author} {\bibfnamefont {C.}~\bibnamefont {Trevillian}}, \bibinfo {author}
  {\bibfnamefont {T.}~\bibnamefont {Polakovic}}, \bibinfo {author}
  {\bibfnamefont {T.~W.}\ \bibnamefont {Cecil}}, \bibinfo {author}
  {\bibfnamefont {P.~S.}\ \bibnamefont {Barry}}, \bibinfo {author}
  {\bibfnamefont {J.}~\bibnamefont {Pearson}}, \bibinfo {author} {\bibfnamefont
  {R.}~\bibnamefont {Divan}}, \bibinfo {author} {\bibfnamefont
  {V.}~\bibnamefont {Tyberkevych}}, \bibinfo {author} {\bibfnamefont {C.~L.}\
  \bibnamefont {Chang}}, \bibinfo {author} {\bibfnamefont {U.}~\bibnamefont
  {Welp}}, \bibinfo {author} {\bibfnamefont {W.-K.}\ \bibnamefont {Kwok}},\
  and\ \bibinfo {author} {\bibfnamefont {V.}~\bibnamefont {Novosad}},\
  }\bibfield  {title} {\bibinfo {title} {Coherent coupling of two remote
  magnonic resonators mediated by superconducting circuits},\ }\href
  {https://doi.org/10.1103/PhysRevLett.128.047701} {\bibfield  {journal}
  {\bibinfo  {journal} {Phys. Rev. Lett.}\ }\textbf {\bibinfo {volume} {128}},\
  \bibinfo {pages} {047701} (\bibinfo {year} {2022}{\natexlab{b}})}\BibitemShut
  {NoStop}%
\bibitem [{\citenamefont {Liu}\ and\ \citenamefont {Houck}(2017)}]{Liu2017}%
  \BibitemOpen
  \bibfield  {author} {\bibinfo {author} {\bibfnamefont {Y.}~\bibnamefont
  {Liu}}\ and\ \bibinfo {author} {\bibfnamefont {A.~A.}\ \bibnamefont
  {Houck}},\ }\bibfield  {title} {\bibinfo {title} {Quantum electrodynamics
  near a photonic bandgap},\ }\href {https://doi.org/10.1038/nphys3834}
  {\bibfield  {journal} {\bibinfo  {journal} {Nat. Phys.}\ }\textbf {\bibinfo
  {volume} {13}},\ \bibinfo {pages} {48} (\bibinfo {year} {2017})}\BibitemShut
  {NoStop}%
\bibitem [{\citenamefont {Krinner}\ \emph {et~al.}(2018)\citenamefont
  {Krinner}, \citenamefont {Stewart}, \citenamefont {Pazmiño}, \citenamefont
  {Kwon},\ and\ \citenamefont {Schneble}}]{Krinner2018}%
  \BibitemOpen
  \bibfield  {author} {\bibinfo {author} {\bibfnamefont {L.}~\bibnamefont
  {Krinner}}, \bibinfo {author} {\bibfnamefont {M.}~\bibnamefont {Stewart}},
  \bibinfo {author} {\bibfnamefont {A.}~\bibnamefont {Pazmiño}}, \bibinfo
  {author} {\bibfnamefont {J.}~\bibnamefont {Kwon}},\ and\ \bibinfo {author}
  {\bibfnamefont {D.}~\bibnamefont {Schneble}},\ }\bibfield  {title} {\bibinfo
  {title} {Spontaneous emission of matter waves from a tunable open quantum
  system},\ }\href {https://doi.org/10.1038/s41586-018-0348-z} {\bibfield
  {journal} {\bibinfo  {journal} {Nature}\ }\textbf {\bibinfo {volume} {559}},\
  \bibinfo {pages} {589} (\bibinfo {year} {2018})}\BibitemShut {NoStop}%
\bibitem [{\citenamefont {Kwon}\ \emph {et~al.}(2022)\citenamefont {Kwon},
  \citenamefont {Kim}, \citenamefont {Lanuza},\ and\ \citenamefont
  {Schneble}}]{Kwon2022}%
  \BibitemOpen
  \bibfield  {author} {\bibinfo {author} {\bibfnamefont {J.}~\bibnamefont
  {Kwon}}, \bibinfo {author} {\bibfnamefont {Y.}~\bibnamefont {Kim}}, \bibinfo
  {author} {\bibfnamefont {A.}~\bibnamefont {Lanuza}},\ and\ \bibinfo {author}
  {\bibfnamefont {D.}~\bibnamefont {Schneble}},\ }\bibfield  {title} {\bibinfo
  {title} {Formation of matter-wave polaritons in an optical lattice},\ }\href
  {https://doi.org/10.1038/s41567-022-01565-4} {\bibfield  {journal} {\bibinfo
  {journal} {Nature Physics}\ }\textbf {\bibinfo {volume} {18}},\ \bibinfo
  {pages} {657} (\bibinfo {year} {2022})}\BibitemShut {NoStop}%
\bibitem [{\citenamefont {Mart\'{\i}nez-Losa~del Rinc\'on}\ \emph
  {et~al.}(2023)\citenamefont {Mart\'{\i}nez-Losa~del Rinc\'on}, \citenamefont
  {Gimeno}, \citenamefont {P\'erez-Bail\'on}, \citenamefont {Rollano},
  \citenamefont {Luis}, \citenamefont {Zueco},\ and\ \citenamefont
  {Mart\'{\i}nez-P\'erez}}]{PhysRevApplied.19.014002}%
  \BibitemOpen
  \bibfield  {author} {\bibinfo {author} {\bibfnamefont {S.}~\bibnamefont
  {Mart\'{\i}nez-Losa~del Rinc\'on}}, \bibinfo {author} {\bibfnamefont
  {I.}~\bibnamefont {Gimeno}}, \bibinfo {author} {\bibfnamefont
  {J.}~\bibnamefont {P\'erez-Bail\'on}}, \bibinfo {author} {\bibfnamefont
  {V.}~\bibnamefont {Rollano}}, \bibinfo {author} {\bibfnamefont
  {F.}~\bibnamefont {Luis}}, \bibinfo {author} {\bibfnamefont {D.}~\bibnamefont
  {Zueco}},\ and\ \bibinfo {author} {\bibfnamefont {M.~J.}\ \bibnamefont
  {Mart\'{\i}nez-P\'erez}},\ }\bibfield  {title} {\bibinfo {title} {Measuring
  the magnon-photon coupling in shaped ferromagnets: Tuning of the resonance
  frequency},\ }\href {https://doi.org/10.1103/PhysRevApplied.19.014002}
  {\bibfield  {journal} {\bibinfo  {journal} {Phys. Rev. Appl.}\ }\textbf
  {\bibinfo {volume} {19}},\ \bibinfo {pages} {014002} (\bibinfo {year}
  {2023})}\BibitemShut {NoStop}%
\bibitem [{\citenamefont {Kansanen}\ \emph {et~al.}(2021)\citenamefont
  {Kansanen}, \citenamefont {Tassi}, \citenamefont {Mishra}, \citenamefont
  {Sillanp\"a\"a},\ and\ \citenamefont {Heikkil\"a}}]{PhysRevB.104.214416}%
  \BibitemOpen
  \bibfield  {author} {\bibinfo {author} {\bibfnamefont {K.~S.~U.}\
  \bibnamefont {Kansanen}}, \bibinfo {author} {\bibfnamefont {C.}~\bibnamefont
  {Tassi}}, \bibinfo {author} {\bibfnamefont {H.}~\bibnamefont {Mishra}},
  \bibinfo {author} {\bibfnamefont {M.~A.}\ \bibnamefont {Sillanp\"a\"a}},\
  and\ \bibinfo {author} {\bibfnamefont {T.~T.}\ \bibnamefont {Heikkil\"a}},\
  }\bibfield  {title} {\bibinfo {title} {Magnomechanics in suspended magnetic
  beams},\ }\href {https://doi.org/10.1103/PhysRevB.104.214416} {\bibfield
  {journal} {\bibinfo  {journal} {Phys. Rev. B}\ }\textbf {\bibinfo {volume}
  {104}},\ \bibinfo {pages} {214416} (\bibinfo {year} {2021})}\BibitemShut
  {NoStop}%
\end{thebibliography}%


\begin{thebibliography}{5}%
\makeatletter
\providecommand \@ifxundefined [1]{%
 \@ifx{#1\undefined}
}%
\providecommand \@ifnum [1]{%
 \ifnum #1\expandafter \@firstoftwo
 \else \expandafter \@secondoftwo
 \fi
}%
\providecommand \@ifx [1]{%
 \ifx #1\expandafter \@firstoftwo
 \else \expandafter \@secondoftwo
 \fi
}%
\providecommand \natexlab [1]{#1}%
\providecommand \enquote  [1]{``#1''}%
\providecommand \bibnamefont  [1]{#1}%
\providecommand \bibfnamefont [1]{#1}%
\providecommand \citenamefont [1]{#1}%
\providecommand \href@noop [0]{\@secondoftwo}%
\providecommand \href [0]{\begingroup \@sanitize@url \@href}%
\providecommand \@href[1]{\@@startlink{#1}\@@href}%
\providecommand \@@href[1]{\endgroup#1\@@endlink}%
\providecommand \@sanitize@url [0]{\catcode `\\12\catcode `\$12\catcode
  `\&12\catcode `\#12\catcode `\^12\catcode `\_12\catcode `\%12\relax}%
\providecommand \@@startlink[1]{}%
\providecommand \@@endlink[0]{}%
\providecommand \url  [0]{\begingroup\@sanitize@url \@url }%
\providecommand \@url [1]{\endgroup\@href {#1}{\urlprefix }}%
\providecommand \urlprefix  [0]{URL }%
\providecommand \Eprint [0]{\href }%
\providecommand \doibase [0]{https://doi.org/}%
\providecommand \selectlanguage [0]{\@gobble}%
\providecommand \bibinfo  [0]{\@secondoftwo}%
\providecommand \bibfield  [0]{\@secondoftwo}%
\providecommand \translation [1]{[#1]}%
\providecommand \BibitemOpen [0]{}%
\providecommand \bibitemStop [0]{}%
\providecommand \bibitemNoStop [0]{.\EOS\space}%
\providecommand \EOS [0]{\spacefactor3000\relax}%
\providecommand \BibitemShut  [1]{\csname bibitem#1\endcsname}%
\let\auto@bib@innerbib\@empty
\bibitem [{\citenamefont {Xiong}\ \emph {et~al.}(2022)\citenamefont {Xiong},
  \citenamefont {Tian}, \citenamefont {Zhang},\ and\ \citenamefont
  {You}}]{PhysRevB.105.245310}%
  \BibitemOpen
  \bibfield  {author} {\bibinfo {author} {\bibfnamefont {W.}~\bibnamefont
  {Xiong}}, \bibinfo {author} {\bibfnamefont {M.}~\bibnamefont {Tian}},
  \bibinfo {author} {\bibfnamefont {G.-Q.}\ \bibnamefont {Zhang}},\ and\
  \bibinfo {author} {\bibfnamefont {J.~Q.}\ \bibnamefont {You}},\ }\bibfield
  {title} {\bibinfo {title} {Strong long-range spin-spin coupling via a {K}err
  magnon interface},\ }\href {https://doi.org/10.1103/PhysRevB.105.245310}
  {\bibfield  {journal} {\bibinfo  {journal} {Phys. Rev. B}\ }\textbf {\bibinfo
  {volume} {105}},\ \bibinfo {pages} {245310} (\bibinfo {year}
  {2022})}\BibitemShut {NoStop}%
\bibitem [{\citenamefont {Neuman}\ \emph {et~al.}(2020)\citenamefont {Neuman},
  \citenamefont {Wang},\ and\ \citenamefont {Narang}}]{PhysRevLett.125.247702}%
  \BibitemOpen
  \bibfield  {author} {\bibinfo {author} {\bibfnamefont {T.}~\bibnamefont
  {Neuman}}, \bibinfo {author} {\bibfnamefont {D.~S.}\ \bibnamefont {Wang}},\
  and\ \bibinfo {author} {\bibfnamefont {P.}~\bibnamefont {Narang}},\
  }\bibfield  {title} {\bibinfo {title} {Nanomagnonic cavities for strong
  spin-magnon coupling and magnon-mediated spin-spin interactions},\ }\href
  {https://doi.org/10.1103/PhysRevLett.125.247702} {\bibfield  {journal}
  {\bibinfo  {journal} {Phys. Rev. Lett.}\ }\textbf {\bibinfo {volume} {125}},\
  \bibinfo {pages} {247702} (\bibinfo {year} {2020})}\BibitemShut {NoStop}%
\bibitem [{\citenamefont {Fletcher}\ and\ \citenamefont
  {Bell}(1959)}]{10.1063/1.1735216}%
  \BibitemOpen
  \bibfield  {author} {\bibinfo {author} {\bibfnamefont {P.~C.}\ \bibnamefont
  {Fletcher}}\ and\ \bibinfo {author} {\bibfnamefont {R.~O.}\ \bibnamefont
  {Bell}},\ }\bibfield  {title} {\bibinfo {title} {Ferrimagnetic resonance
  modes in spheres},\ }\href {https://doi.org/10.1063/1.1735216} {\bibfield
  {journal} {\bibinfo  {journal} {Journal of Applied Physics}\ }\textbf
  {\bibinfo {volume} {30}},\ \bibinfo {pages} {687} (\bibinfo {year}
  {1959})}\BibitemShut {NoStop}%
\bibitem [{\citenamefont {Walker}(1957)}]{PhysRev.105.390}%
  \BibitemOpen
  \bibfield  {author} {\bibinfo {author} {\bibfnamefont {L.~R.}\ \bibnamefont
  {Walker}},\ }\bibfield  {title} {\bibinfo {title} {Magnetostatic modes in
  ferromagnetic resonance},\ }\href {https://doi.org/10.1103/PhysRev.105.390}
  {\bibfield  {journal} {\bibinfo  {journal} {Phys. Rev.}\ }\textbf {\bibinfo
  {volume} {105}},\ \bibinfo {pages} {390} (\bibinfo {year}
  {1957})}\BibitemShut {NoStop}%
\bibitem [{\citenamefont {Walker}(2004)}]{10.1063/1.1723117}%
  \BibitemOpen
  \bibfield  {author} {\bibinfo {author} {\bibfnamefont {L.~R.}\ \bibnamefont
  {Walker}},\ }\bibfield  {title} {\bibinfo {title} {Resonant modes of
  ferromagnetic spheroids},\ }\href {https://doi.org/10.1063/1.1723117 %J
  Journal of Applied Physics} {\bibfield  {journal} {\bibinfo  {journal}
  {Journal of Applied Physics}\ }\textbf {\bibinfo {volume} {29}},\ \bibinfo
  {pages} {318} (\bibinfo {year} {2004})}\BibitemShut {NoStop}%
\end{thebibliography}%
\end{document}


\title{Supplemental Material for ``Kerr nonlinearity induced strong spin-magnon coupling''}
\author{Feng-Zhou Ji\orcid{0000-0003-4859-1535}}
\affiliation{Key Laboratory of Quantum Theory and Applications of MoE, Lanzhou Center for Theoretical Physics, and Key Laboratory of Theoretical Physics of Gansu Province, Lanzhou University, Lanzhou 730000, China}
\author{Jun-Hong An\orcid{0000-0002-3475-0729}}
\email{anjhong@lzu.edu.cn}
\affiliation{Key Laboratory of Quantum Theory and Applications of MoE, Lanzhou Center for Theoretical Physics, and Key Laboratory of Theoretical Physics of Gansu Province, Lanzhou University, Lanzhou 730000, China}

\maketitle

\section{Magnon-spin coupling system within Kerr nonlinearity}
We consider a spin defect as a magnetic emitter coupled to the magnons attached to a YIG sphere in the presence of the Kerr nonlinearity. The Hamiltonian reads
\begin{eqnarray}
\hat{\mathcal{H}}_{\text{Kerr}}&=&\hbar\omega_{0}\hat{\sigma}^{\dagger}\hat{\sigma}+\sum_{k}[\hbar\omega_{k}\hat{b}_{k}^{\dagger}\hat{b}_{k}-(\hbar K/2)\hat{b}_{k}^{\dagger2}\hat{b}_{k}^{2}\nonumber\\
&&-(g_{k}\hat{b}_{k}^{\dagger}\hat{\sigma}-\Omega_{d}e^{i\phi_d}\hat{b}_{k}^{\dagger}e^{-i\omega_{d}t}+\text{H.c.})],\label{smhmt}
\end{eqnarray}
where $\hat{\sigma}=|g\rangle\langle e|$ is the transition operator of the spin defect with frequency $\omega_0$, $\hat{b}_k$ is the annihilation operator of $k$-th magnon with frequency $\omega_k$. The  $g_{k}=\mu_{0}\mathbf{m}\cdot\tilde{\mathbf{H}}_{k}^{\ast}(\mathbf{r})$ is related to the vacuum amplitude of $k$th magnon mode. In the rotating frame with $\hat{H}_0=\omega_{d}(\sum_{k}\hat{b}_{k}^{\dagger}\hat{b}_{k}+\hat{\sigma}^{\dagger}\hat{\sigma})$,
Eq. \eqref{smhmt} becomes
\begin{eqnarray}
\hat{\mathcal{H}}_{1}&=&\hbar\Delta_{0}\hat{\sigma}^{\dagger}\hat{\sigma}+\sum_{k}[\hbar(\omega_{k}-\omega_{d})\hat{b}_{k}^{\dagger}\hat{b}_{k}-(\hbar K/2)\hat{b}_{k}^{\dagger2}\hat{b}_{k}^{2}\nonumber \\
&&-(g_{k}\hat{b}_{k}^{\dagger}\hat{\sigma}-\Omega_{d}e^{i\phi_d}\hat{b}_{k}^{\dagger}+\text{H.c.})],
\end{eqnarray}
where $\Delta_0=\omega_0-\omega_d$. The Heisenberg-Langevin equation satisfied by $\hat{b}_k$ is
\begin{eqnarray}
\frac{d\hat{b}_{k}}{dt}	&=&-i[(\omega_{k}-\omega_{d})\hat{b}_{k}-K\hat{b}_{k}^{\dagger}\hat{b}_{k}^2-{g_{k}\over\hbar}\hat{\sigma}+{\Omega_{d}e^{i\phi_d}\over\hbar}]\nonumber\\
&&-\Gamma_0\hat{b}_{k}+\sqrt{2\Gamma_0}\hat{\mathcal F}_{noise},
\end{eqnarray}
where $\Gamma_0$ is the damping rate of magnon, $\hat{\mathcal F}_{noise}$ is the vacuum noise with a zero expectation value. Decomposing $\hat{b}_k$ as its steady-state expectation value and fluctuation, i.e., $\hat{b}_{k}=\langle\hat{b}_{k}\rangle+\delta\hat{b}_{k}$, we have
\begin{eqnarray}
0 =-i[(\omega_{k}-\omega_{d})\langle\hat{b}_{k}\rangle+{\Omega_{d}e^{i\phi_d}\over\hbar}- KN_{k}\langle\hat{b}_{k}\rangle]-\Gamma_0\langle\hat{b}_{k}\rangle,~~
\end{eqnarray}where we have used the mean-field approximation by replacing $\hat{b}_k^\dag\hat{b}_k$ as its expectation value $N_k=\langle \hat{b}_k^\dag\hat{b}_k\rangle\simeq |\langle\hat{b}_k\rangle|^2$ under the large-$N_k$ condition. We obtain
\begin{equation}
\langle\hat{b}_{k}\rangle=\frac{-\Omega_{d}e^{i\phi_d}/\hbar}{(\omega_{k}-\omega_{d}-KN_{k}-i\Gamma_0)}\simeq \frac{-\Omega_{d}e^{i\phi_d}/\hbar}{(\omega_{k}-\omega_{d}-i\Gamma_0)}.\label{smdfd1}
\end{equation}under the condition $\omega_d\gg K$. The large-$N_k$ condition required by the mean-field approximation is valid in the regime with a large $\Omega_d$ and $\omega_d\gg K$. Then we can use the large $\langle\hat{b}_k\rangle$ to linearize the Hamiltonian. The equations of motion of the fluctuation operators becomes
\begin{eqnarray}
\frac{d\delta\hat{b}_{k}}{dt} & =&-i[(\omega_{k}-\omega_{d}-2KN_k)\delta\hat{b}_{k}-{g_{k}\over\hbar}\hat{\sigma}]+\sqrt{2\Gamma_0}\hat{\mathcal F}_{noise}\nonumber\\
 &&-\Gamma_0\delta\hat{b}_{k} +iK\langle\hat{b}_{k}\rangle^{2}\delta\hat{b}_{k}^{\dagger},\label{smlinds}
\end{eqnarray}where the higher-order terms of the fluctuation operators have been neglected \cite{PhysRevB.105.245310}. By properly tuning the phase $\phi_d$ of the driving field, we may obtain a real $\langle\hat{b}_k\rangle$. Then $\langle \hat{b}_k\rangle^2=N_k$. Equation \eqref{smlinds} describes a dynamics governed by a linearized Hamiltonian
\begin{eqnarray}
\hat{\mathcal{H}}_{\text{Linear}}&=&\hbar\Delta_{0}\hat{\sigma}^{\dagger}\hat{\sigma}+\sum_{k}[(\hbar(\omega_{k}-\Pi_k)\delta\hat{b}_{k}^{\dagger}\delta\hat{b}_{k}\nonumber \\
&&-(g_{k}\delta\hat{b}_{k}^{\dagger}\hat{\sigma}+(\hbar \mathcal{K}_k/ 2)\delta\hat{b}_{k}^{2}+\text{H.c.})],\label{smlnd}
\end{eqnarray}where $\Pi_k=\omega_d+2K\langle\hat{b}_k\rangle^2$ and $\mathcal{K}_k=K\langle\hat{b}_k\rangle^2$. A Bogoliubov transformation $\hat{\mathcal S}=\prod_ke^{{r_k\over 2}(\delta \hat{b}_k^2-\delta\hat{b}_k^{\dag 2})}$, with $r_k=\frac{1}{4}\ln(\frac{\omega_{k}-\Pi_k+\mathcal{K}_k}{\omega_{k}-\Pi_k-\mathcal{K}_k})$, converts Eq. \eqref{smlnd} into
\begin{equation}
\hat{\mathcal{H}}=\hbar\Delta_{0}\hat{\sigma}^{\dagger}\hat{\sigma}+\sum_{k}[(\hbar\zeta_{k}\delta\hat{b}_{k}^{\dagger}\delta\hat{b}_{k}-(\mathcal{G}_{k}\delta\hat{b}_{k}^{\dagger}\hat{\sigma}+\text{H.c.})],
\end{equation}under the rotating-wave approximation, where $\zeta_k=(\omega_k-\Pi_k)/\cosh(2r_k)$ and $\mathcal{G}_k=g_ke^{r_k}/2$. After replacing $\delta\hat{b}_k$ by $\hat{b}_k$ for brevity, we arrive at the linearized Hamiltonian in Eq. (3) of the main text.

We can tune $\omega_d$ such that $r_k$ and $\mathcal{K}_k$ exhibit a soft dependence on $k$. We can use a relative large $\omega_d$ such that $\omega_d\gg\mathcal{K}_k=K|\langle \hat{b}_{k}\rangle|^2$. It makes us roughly approximate $\mathcal{K}_k\simeq \mathcal{K}$, which reduces $r_k$ into
$r_{k}\simeq \frac{1}{4}\ln(\frac{\omega_{k}-\omega_{d}-\mathcal{K}}{\omega_{k}-\omega_{d}-3\mathcal{K}})$.
Further, because $\omega_k$ is within a finite band regime, for example, $\omega_k$ of the YIG is limited between $14$ GHz to $16.5$ GHz \cite{PhysRevLett.125.247702}, we can neglect the $k$-dependence of $r_k$ too. Thus, we phenomenologically investigate the effect of the Kerr nonlinearity induced by a constant $r$ for simplification.

\section{Spectral density}

We derive the spectral density using the Green's tensor method. It can be readily derived from Eq. (3) in the main text that the spectral density reads
\begin{align}
J(\zeta)=&\sum_k{|\mathcal{G}_k|^2\over\hbar^2}\delta(\zeta-\zeta_k)\nonumber\\
=&\frac{e^{2r}\mu_{0}^{2}}{4\hbar^2}\textbf{m}^{\ast}\cdot\sum_{k}|\tilde{\textbf{H}}_{k}(\textbf{r})|^2\delta(\zeta-\zeta_{k})\cdot\textbf{m}.\label{spectrumfulu1}
\end{align}
In order to derive $\tilde{\textbf{H}}_{k}(\textbf{r})$, we investigate a magnetic field induced by a magnetic dipole $\bar{{\bf m}}$ at $\textbf{a}$ and oscillating at frequency $\bar{\omega}$. The magnetic-field strength at ${\bf r}$ can be represented by the Green's tensor as
\begin{equation}
\textbf{H}(\textbf{r})=\bar{k}^2\textbf{G}(\textbf{r},\textbf{a},\bar{\omega})\cdot\bar{\textbf{m}},\label{grf}
\end{equation}
where $\bar{k}=\bar{\omega}/c$. 
The process is semiclassically described by
\begin{equation}
\hat{H}_{\text{source}}=\sum_k[\hbar\omega_k\hat{b}_{k}^{\dagger}\hat{b}_{k}-(g_{k}e^{-i\bar{\omega}t}\hat{b}_{k}^{\dagger}+\text{H.c.})],\label{cizi3}
\end{equation}
where $g_k=\mu_{0}\bar{\textbf{m}}\cdot\tilde{\textbf{H}}_{k}^{\ast}(\textbf{a})$.
According to the Heisenberg equation satisfied by $\hat{b}_{k}$, i.e.,
\begin{equation}
\frac{d\hat{b}_{k}}{dt}=-i\omega_{k}\hat{b}_{k}+i\frac{g_{k}}{\hbar}e^{-i\bar{\omega}t}.
\end{equation}
and the initial condition $\langle \hat{b}_{k}(0)\rangle=0$, we get $\langle\hat{b}_{k}(t)\rangle=i\frac{g_{k}}{\hbar}\int_{0}^{t}d\tau e^{-i(\bar{\omega}-\omega_{k})\tau}$. It long-time limit reads
\begin{equation}
\langle\hat{b}_{k}(\infty)\rangle=\frac{g_{k}}{\hbar}\frac{1}{\bar{\omega}-\omega_{k}-i0^{+}}.
\end{equation}
Then the magnetic field is recast into
\begin{eqnarray}
\textbf{H}(\textbf{r})=\sum_{k}\tilde{\textbf{H}}_{k}(\textbf{r})\langle \hat{b}_{k}(\infty)\rangle={\mu_0\over\hbar}\sum_{k}\frac{\tilde{\textbf{H}}_{k}(\textbf{r})\tilde{\bf H}_{k}^{\ast}(\textbf{a})\cdot\bar{\textbf{m}}}{\bar{\omega}-\omega_{k}-i0^{+}}.~~\label{cmag}
\end{eqnarray}
Comparing Eq. \eqref{grf} with Eq. \eqref{cmag}, we have
\begin{equation}
\bar{k}^{2}\textbf{G}(\textbf{r},\textbf{a},\bar{\omega})={\mu_0\over\hbar}\sum_{k}\frac{\tilde{\textbf{H}}_{k}(\textbf{r})\tilde{\textbf{H}}_{k}^{\ast}(\textbf{a})}{\bar{\omega}-\omega_{k}-i0^{+}}.\label{cizi5}
\end{equation}
Using the identity $\frac{1}{\bar{\omega}-\zeta_{k}-i0^{+}}=i\pi\delta(\bar{\omega}-\omega_{k})+\mathcal{P}\frac{1}{\bar{\omega}-\omega_{k}}$, Eq. \eqref{cizi5} results in  $\text{Im}[\bar{k}^{2}\textbf{G}(\textbf{r},\textbf{a},\bar{\omega})]=\frac{\pi\mu_0}{\hbar}\sum_{k}\tilde{\textbf{H}}_{k}(\textbf{r})\tilde{\textbf{H}}_{k}^{\ast}(\textbf{a})\delta(\bar{\omega}-\omega_{k})$. Substituting $\zeta_{k}=(\omega_{k}-\Pi)/\cosh{(2r)}$, with $\Pi=\omega_{d}$, into Eq. \eqref{spectrumfulu1}, we obtain
\begin{eqnarray}
J(\zeta)&=&\frac{\eta\mu_{0}^2}{4\hbar^2}\textbf{m}^{\ast}\cdot\sum_{k}|\tilde{\textbf{H}}_{k}(\textbf{r})|^2\delta(\cosh(2r)\zeta+\Pi-\omega_k)\cdot\textbf{m}\nonumber\\
&=&\frac{\eta\mu_{0}}{4\hbar\pi}\text{Im}[\textbf{m}^{\ast}\cdot k^2\textbf{G}(\textbf{r},\textbf{r},\zeta\cosh(2r)+\Pi)\cdot\textbf{m}],\label{spectrumfulu2}
\end{eqnarray}
where $k=[\zeta\cosh(2r)+\Pi]/c$ and $\eta=e^{2r}\cosh(2r)$.

\section{Magnetostatic Green's Tensor}
We provide the detailed derivation of the magnetostatic Green's tensor induced by a point dipole near the YIG sphere \cite{PhysRevLett.125.247702,10.1063/1.1735216}.
The Green's tensor is given by
\begin{eqnarray}
\bar{k}^{2}\textbf{G}(\textbf{r}, \textbf{a},\bar{\omega})&=&\bar{k}^{2}\textbf{G}_{0}(\textbf{r},\textbf{a},\bar{\omega})+\sum_{\alpha,\beta\in \{r,\theta,\varphi\}}H_{\alpha}^{\beta}\textbf{e}_{\alpha}\textbf{e}_{\beta}\nonumber\\
&=&\sum_{\alpha,\beta\in \{r,\theta,\varphi\}}[H_{0,\alpha}^{\beta}+H_{\alpha}^{\beta}]\textbf{e}_{\alpha}\textbf{e}_{\beta},\label{greentensorfulu1}
\end{eqnarray}
where $\textbf{G}_{0}(\textbf{r},\textbf{a},\bar{\omega})$ is the Green's tensor describing the magnetic field of a magnetic dipole in a vacuum. The magnetic fields ${\bf H}=-{\pmb \nabla}\phi$ and $\phi={1\over 4\pi}{\pmb \nabla}_{ a}{1\over |{\bf r}-{\bf a}|}$. $\bf{H}_0$ takes the similar form as ${\bf H}$ but in the absence of the YIG sphere.

The potential of a point charge can be expressed in standard spherical coordinates as \cite{PhysRevLett.125.247702}
\begin{equation}
\frac{1}{|\textbf{r}-\textbf{a}|}=\sum_{lm}\frac{(l-m)!}{(l+m)!}R_{l,m}(r,\theta,\varphi)I_{l,m}^{\ast}(a,\theta_{a},\varphi_{a}),\label{xxxx}
\end{equation}
where $I_{n,m}(\textbf{R})$ and $R_{n,m}(\textbf{R})$ are the solid harmonics:
\begin{equation}
\begin{aligned}
I_{n,m}(\textbf{R})&=\mathcal{R}^{-n-1}P_{n}^{m}(\cos\Theta)e^{im\Phi}\\
R_{n,m}(\textbf{R})&=\mathcal{R}^{n}P_{n}^{m}(\cos\Theta)e^{im\Phi}.
\end{aligned}
\end{equation}
The potential of the dipole can be obtained by calculating the derivative of Eq. \eqref{xxxx} with respect to the coordinates $\{a,\theta_a,\varphi_a\}$ of the source. The results are
\begin{equation}
\phi_{{\alpha}}=\sum_{lm}D_{lm}^{\alpha}r^{l}P_{l}^{m}(\cos\theta)e^{im\varphi},\label{apppotd}
\end{equation}
where $\alpha\in(r,\theta,\varphi)$ and
\begin{equation}
\begin{aligned}
D_{lm}^{r}&=-\frac{1}{4\pi}\frac{(l-m)!}{(l+m)!}\frac{(l+1)}{a^{l+2}}P_{l}^{m}(\cos\theta_{a})e^{-im\varphi_{a}},\\
D_{lm}^{\theta}&=-\frac{1}{4\pi}\frac{(l-m)!}{(l+m)!}\frac{\sin\theta_{a}}{a^{l+2}}P_{l}^{m\prime}(\cos\theta_{a})e^{-im\varphi_{a}},\\
D_{lm}^{\varphi}&=-\frac{1}{4\pi}\frac{(l-m)!}{(l+m)!}\frac{im}{a^{l+2}\sin \theta_{a}}P_{l}^{m}(\cos\theta_{a})e^{-im\varphi_{a}}.
\end{aligned}
\end{equation}

In the presence of the YIG sphere, the boundary condition should be carefully considered. The equation of motion of the potential $\phi$ is
\begin{eqnarray}
\left\{
\begin{array}{cc}
(1+{\chi})(\frac{\partial^{2}}{\partial x^{2}}+\frac{\partial^{2}}{\partial y^{2}})\phi+\frac{\partial^{2}\phi}{\partial z^{2}}=0 , & r\leq R \\
(\frac{\partial^{2}}{\partial x^{2}}+\frac{\partial^{2}}{\partial y^{2}}+\frac{\partial^{2}}{\partial z^{2}})\phi=0 , & r>R
\end{array},
\right.\label{appdphi}
\end{eqnarray}
where $\chi$ is magnetic susceptibility tensor. ${\pmb\chi}$ can be calculated from the Landau-Lifschitz-Gilbert equation \cite{PhysRev.105.390,10.1063/1.1735216} as
\begin{equation}
\begin{aligned}
\chi_{xx}&=\chi_{yy}=\frac{\gamma^{2}h_{0}M_{s}}{\gamma^{2}h_{0}^{2}-\omega^{2}-i\Gamma_{0}\omega}\equiv\chi,\\
\chi_{xy}&=\chi^{\ast}_{yx}=i\frac{\gamma\omega M_{s}}{\gamma^{2}h_{0}^{2}-\omega^{2}-i\Gamma_{0}\omega}\equiv i\kappa,\label{magsus}
\end{aligned}
\end{equation}
where $\gamma$ is the gyromagnetic ratio, $\Gamma_{0}= 2\gamma h_{0}\alpha$ is the damping parameter, and $M_{s}$ is the saturation magnetization. Here, the static magnetic field $\textbf{h}_{0}=h_{0}\textbf{e}_{z}=\textbf{H}_{e}+\textbf{H}_{d}$, where $\textbf{H}_{e}$ is the external static field and $\textbf{H}_{d}=-\textbf{M}_{s}/3$ is the demagnetization field.

Introducing the ellipsoidal coordinates,
\begin{equation}
\begin{aligned}
x&=R\sqrt{-\chi}\sqrt{1-\xi^{2}}\sin\eta\cos\varphi,\\
y&=R\sqrt{-\chi}\sqrt{1-\xi^{2}}\sin\eta\sin\varphi,\\
z&=R\sqrt{\frac{\chi}{1+\chi}}\xi \cos\eta,
\end{aligned}
\end{equation}
the general solution of Eq. \eqref{appdphi} is \cite{PhysRevLett.125.247702,10.1063/1.1735216}
\begin{eqnarray}
\phi_{\alpha}^\text{in}&=&\sum_{nm}A_{nm}^{\alpha}P_{n}^{m}(\xi)P_{n}^{m}(\cos\eta)e^{im\varphi}, ~r\leq R, \\
\phi_\alpha^\text{out}&=&\sum_{nm}[{B_{nm}^{\alpha}\over r^{n+1}}+D_{nm}^{\alpha}r^{n}]P_{n}^{m}(\cos\theta)e^{im\varphi}, ~ r>R.~~~~~~~
\end{eqnarray}
They satisfy the boundary conditions $(\phi_\alpha^{\text{in}})_{R}=(\phi_\alpha^{\text{out}})_{R}$ and
\begin{equation}
\begin{aligned}
(\frac{\partial \phi_\text{out}}{\partial r})_{R}&=(1+\chi \sin^{2}\theta)(\frac{\partial\phi_\text{in}}{\partial r})_{R}\\
&+\frac{\chi\sin\theta\cos\theta}{R}(\frac{\partial\phi_\text{in}}{\partial\theta})_{R}+\frac{i\kappa}{R}(\frac{\partial\phi_\text{in}}{\partial\varphi})_{R}.
\end{aligned}
\end{equation}
Noticing $\eta=\theta$ and $\xi_{0}^{2}=1+1/\chi$ at the sphere surface, we obtain $B_{nm}^{\alpha}$ represented by $D_{nm}^{\alpha}$ as
\begin{equation}
B_{nm}^{\alpha}=\frac{D_{nm}^{\alpha}R^{2n+1}[(n+\kappa m)P_{n}^{m}(\xi_{0})-\xi_{0}P_{n}^{m\prime}(\xi_{0})]}{(n+1-\kappa m)P_{n}^{m}(\xi_{0})+\xi_{0}P_{n}^{m\prime}(\xi_{0})}.\label{smbnm}
\end{equation}
We separate $\phi_\alpha^\text{out}$ into the free-vacuum and the YIG-induced contributions as
\begin{eqnarray}
\phi_\alpha^\text{out}&=&\phi_\alpha^\text{(0)}+\phi_\alpha^\text{ind},\\
\phi_\alpha^\text{(0)}&=&\sum_{nm}D_{nm}^{\alpha}r^{n}P_{n}^{m}(\cos\theta)e^{im\varphi},\\
\phi_\alpha^\text{ind}&=&\sum_{nm}\frac{D_{nm}^{\alpha}R^{2n+1}[(n+\kappa m)P_{n}^{m}(\xi_{0})-\xi_{0}P_{n}^{m\prime}(\xi_{0})]}{(n+1-\kappa m)P_{n}^{m}(\xi_{0})+\xi_{0}P_{n}^{m\prime}(\xi_{0})}\nonumber\\
&&\times r^{-n-1}P_{n}^{m}(\cos\theta)e^{im\varphi}
\end{eqnarray}
Then the free-vacuum contributed magnetic field can be calculated via $\textbf{H}_{0}=-{\pmb \nabla}\phi_\alpha^{(0)}$ as
\begin{align}
H_{0,r}^{\alpha} & =-\sum_{nm}D_{nm}^{\alpha}nr^{n-1}P_{n}^{m}(\cos\theta)e^{im\varphi},\label{apph0r}\\
H_{0,\theta}^{\alpha} & =\sum_{nm}D_{nm}^{\alpha}r^{n-1}P_{n}^{m\prime}(\cos\theta)\sin\theta e^{im\varphi},\label{apph0tht}\\
H_{0,\varphi}^{\alpha} & =-\sum_{nm}D_{nm}^{\alpha}r^{n-1}P_{n}^{m}(\cos\theta)\sin\theta^{-1}ime^{im\varphi}.\label{apph0varp}
\end{align}
The YIG-induced magnetic filed is calculated via $\bf{H}=-{\pmb \nabla}\phi_\alpha^\text{ind}$ as
\begin{eqnarray}
H_{r}^{\alpha}&=&\sum_{nm}\frac{D_{nm}^{\alpha}R^{2n+1}[(n+\kappa m)P_{n}^{m}(\xi_{0})-\xi_{0}P_{n}^{m\prime}(\xi_{0})]}{(n+1-\kappa m)P_{n}^{m}(\xi_{0})+\xi_{0}P_{n}^{m\prime}(\xi_{0})}\nonumber\\
&&\times (n+1)\frac{P_{n}^{m}(\cos\theta)e^{im\varphi}}{r^{n+2}},\label{apphr}\\
H_{\theta}^{\alpha}&=&\sum_{nm}\frac{D_{nm}^{\alpha}R^{2n+1}[(n+\kappa m)P_{n}^{m}(\xi_{0})-\xi_{0}P_{n}^{m\prime}(\xi_{0})]}{(n+1-\kappa m)P_{n}^{m}(\xi_{0})+\xi_{0}P_{n}^{m\prime}(\xi_{0})}\nonumber\\
&&\times \frac{\sin\theta P_{n}^{m\prime}(\cos\theta)e^{im\varphi}}{r^{n+2}},\label{apphtht}\\
H_{\varphi}^{\alpha}&=&\sum_{nm}\frac{D_{nm}^{\alpha}R^{2n+1}[(n+\kappa m)P_{n}^{m}(\xi_{0})-\xi_{0}P_{n}^{m\prime}(\xi_{0})]}{(n+1-\kappa m)P_{n}^{m}(\xi_{0})+\xi_{0}P_{n}^{m\prime}(\xi_{0})}\nonumber\\
&&\times \frac{-imP_{n}^{m}(\cos\theta)e^{im\varphi}}{r^{n+2}\sin\theta}.\label{apphvarp}
\end{eqnarray}
Substituting Eqs. \eqref{apph0r}, \eqref{apph0tht}, \eqref{apph0varp} and Eqs. \eqref{apphr}, \eqref{apphtht}, \eqref{apphvarp} into Eq. \eqref{greentensorfulu1}, we finally obtain the Green's tensor.

We see from Eq. \eqref{smbnm} that the resonance occurs when
\begin{equation}
(n+1-m\kappa)P_{n}^{m}(\xi_0)+\xi_{0}P_{n}^{m{\prime}}(\xi_{0})=0. \label{respseculareq}
\end{equation}The magnon modes corresponding to $n=-m=1$, $2$, and $3$ in the absence of the Kerr nonlinearity and the magnon damping, i.e., $\Gamma_0=0$, are the dipole or Kittel mode $\omega_{\text{K}}=\gamma(h_{0}+M_s/3)$, the quadrupolar mode $\omega_{\text{Q}}=\gamma(h_{0}+2M_s/5)$, and the octupolar mode $\omega_{\text{O}}=\gamma(h_{0}+3M_s/7)$, respectively. Rewriting Eq. \eqref{magsus} for $\Gamma_0=0$ as
\begin{equation}
\begin{aligned}
\chi_{xx}&=\chi_{yy}=\frac{\Omega_H}{\Omega_H^2-\Omega^2}\equiv\chi,\\
\chi_{xy}&=\chi^{\ast}_{yx}=i\frac{\Omega}{\Omega_H^2-\Omega^2}\equiv i\kappa,\label{respmagsus}
\end{aligned}
\end{equation}
with $\Omega_H={h_0\over M_s}$ and $\Omega={\omega\over \gamma M_s}$. Walker proved a significant result that there is a relation $0\le(\Omega-\Omega_{H})\le 1/2$ when $n$ and $m$ ranges all their permitted values \cite{10.1063/1.1723117}. Therefore, the magnonic frequency regime is $\gamma h_0\le\omega\le \gamma(h_0+M_s/2)$, which is finite in its bandwidth.
In our manuscript, the Kerr nonlinearity renormalizes the the magnon frequency as $(\gamma h_0-\Pi)/\cosh(2r)\le\zeta\le (\gamma(h_0+M_s/2)-\Pi)/\cosh(2r)$. This finite bandwidth is verified by our numerical result of spectral density obtained via calculating the the Green's tensor, see Fig. 2(a). Thus, the integral range of the spectral density is limited by $\zeta_{\text{min}}=(\gamma h_0-\Pi)/\cosh(2r)$ and $\zeta_{\text{max}}=(\gamma(h_0+M_s/2)-\Pi)/\cosh(2r)$. The presence of the magnon damping $\Gamma_0$ only results in the expansion of the resonance peaks in the spectral density.
\section{Energy spectrum}
The eigen state of $\hat{\mathcal H}$ in the single-excitation subspace can be expanded as $|\phi_{E}\rangle=x|e,\{0_k\}\rangle+\sum_{k}y_{k}|g,1_{k}\rangle$. According to $\hat{\mathcal H}|\phi_{E}\rangle=E|\phi_{E}\rangle$, we obtain
\begin{eqnarray}
Ex&=&x\hbar\Delta_0-\sum_{k}y_{k}\mathcal{G}_{k}^{\ast}\\
Ey_{k}&=&y_{k}\hbar\zeta_{k}-x\mathcal{G}_{k}.
\end{eqnarray}
Combining these two equations, we have
\begin{equation}
E-\hbar\Delta_{0}=\sum_{k}\frac{|\mathcal{G}_{k}|^{2}}{E-\hbar\zeta_{k}}.\label{smsdd}
\end{equation}
Substituting $J(\zeta)=\sum_{k}\frac{|\mathcal{G}_{k}|^{2}}{\hbar^{2}}\delta(\zeta-\zeta_{k})$ into Eq. \eqref{smsdd} in the continuous limit of the magnonic frequencies, we obtain
\begin{equation}
E/\hbar=\Delta_{0}+\int_{\zeta_\text{min}}^{\zeta_\text{max}}\frac{J(\zeta)}{E/\hbar-\zeta}d\zeta,
\end{equation}which matches with the pole equation (10) in the main text.

\section{Dynamics and steady state}
Consider that the YIG sphere is at low temperature such that the magnons are initially in the vacuum state. To derive the non-Markovian master equation, we consider the following special case of the initial state of the spin.
\begin{enumerate}
    \item The initial state is $|\Psi_1(0)\rangle=|g,\{0_k\}\rangle$. It is easy to check that, governed by Eq. (3) in the main text, $|\Psi_1(t)\rangle=|\Psi(0)\rangle$.
    \item The initial state is $|\Psi_2(0)\rangle=|e,\{0_k\}$. Its time evolution goverend by Eq. (3) is expanded as
    \begin{equation}
        |\Psi_2(t)\rangle=c(t)|e,\{0_k\}\rangle+\sum_k d_k(t)|g,1_k\rangle.
    \end{equation}From $i\hbar|\dot{\Psi}(t)\rangle=\mathcal{H}|\Psi(t)\rangle$, we derive
\begin{eqnarray}
    i\dot{c}(t)&=&\Delta_0c(t)-\sum_k{\mathcal{G}_k^{\ast}\over\hbar} d_k(t),\label{smcdtd}\\
    i\dot{d}_k(t)&=&\zeta_k d_k(t)-{\mathcal{G}_k\over\hbar} c(t).
\end{eqnarray}Substituting the solution $d_k(t)=i{\mathcal{G}_k\over\hbar}\int_0^te^{-i\zeta_k(t-\tau)}c(\tau)d\tau$ into Eq. \eqref{smcdtd}, we obtain
\begin{equation}
    \dot{c}(t)+i\Delta_0c(t)+\int_0^tf(t-\tau)c(\tau)d\tau, \label{smct}
\end{equation}under $c(0)=1$, where we have defined $J(\zeta)=\sum_k{|\mathcal{G}_k|^2\over\hbar^2}\delta(\zeta-\zeta_k)$ and $f(t-\tau)=\int_{\zeta_\text{min}}^{\zeta_\text{max}}d\zeta J(\zeta)e^{-i\zeta(t-\tau)}$.
\end{enumerate}
With these two special cases at hand, the evolution of an arbitrary initial state of the spin $\rho_\text{tot}(0)=\rho(0)\otimes |\{0_k\}\rangle\langle \{0_k\}|$, where $\rho(0)=\rho_{ee}|e\rangle\langle e|+\rho_{gg}|g\rangle\langle g| +(\rho_{eg}|e\rangle\langle g|+\text{h.c.})$ reads
\begin{eqnarray}
    \rho_\text{tot}(t)&=&\rho_{ee}|\Psi_2(t)\rangle\langle\Psi_2(t)|+\rho_{gg}|g,\{0_k\}\rangle\langle g,\{0_k\}|
    \nonumber\\&&+[\rho_{eg}|\Psi_2(t)\rangle\langle g,\{0_k\}|+\text{h.c.}].
\end{eqnarray}
After tracing over the magnonic degrees of freedom, we obtain
\begin{eqnarray}
    \rho(t)&=&\text{Tr}_m[\rho_\text{tot}(t)]\nonumber\\
    &=&\rho_{ee}|c(t)|^2|e\rangle\langle e|+(1-\rho_{ee}|c(t)|^2)|g\rangle\langle g|\nonumber\\
    &&+[\rho_{eg}c(t)|e\rangle\langle g|+\text{h.c.}],\label{smrdcd}
\end{eqnarray}where $\rho_{ee}+\rho_{gg}=1$ has been used. In the basis formed by $|e\rangle$ and $|g\rangle$, its time derivative is
\begin{eqnarray}
\dot{\rho}(t)	&=&\Gamma(t)\left(\begin{array}{cc}
-2\rho_{ee}|c(t)|^{2} & -\rho_{eg}c(t)\\
-\rho_{ge}c^{\ast}(t) & 2\rho_{ee}|c(t)|^{2}
\end{array}\right)\nonumber\\
&&-i\Omega(t)\left(\begin{array}{cc}
0 & \rho_{eg}c(t)\\
-\rho_{ge}c^{\ast}(t) & 0
\end{array}\right),\label{densitymatrix2}
\end{eqnarray}
where $\Gamma(t)=-\text{Re}[\frac{\dot{c}(t)}{c(t)}]$ and $\Omega(t)=-\text{Im}[\frac{\dot{c}(t)}{c(t)}]$.
Rewriting $\left(\begin{array}{cc}
0 & \rho_{eg}c(t)\\
-\rho_{ge}c^{\ast}(t) & 0
\end{array}\right)=	[\hat{\sigma}^{\dagger}\hat{\sigma},\rho(t)]$ and $\left(\begin{array}{cc}
-2\rho_{ee}|c(t)|^{2} & -\rho_{eg}c(t)\\
-\rho_{ge}c^{\ast}(t) & 2\rho_{ee}|c(t)|^{2}\end{array}\right)=2\hat{\sigma}\rho(t)\hat{\sigma}^{\dagger}-\hat{\sigma}^{\dagger}\hat{\sigma}\rho(t)-\rho(t)\hat{\sigma}^{\dagger}\hat{\sigma} $, we finally obtain the exact master equation of the spin as
\begin{equation}
\dot{\rho}(t)=i\Omega(t)[\rho(t),\hat{\sigma}^\dag\hat{\sigma}]+\Gamma(t)[2\hat{\sigma}\rho(t)\hat{\sigma}^\dag
-\{\hat{\sigma}^\dag\hat{\sigma},\rho(t)\}].\label{smexactME}
\end{equation}In our investigation, the initial state of the spin is $\rho(0)=|e\rangle\langle e|$. Thus, from Eq. \eqref{smrdcd}, we have $\rho(t)=|c(t)|^2|e\rangle\langle e|+[1-|c(t)|^2]|g\rangle\langle g|$. The excited state population of the spin is just $|c(t)|^2$.

The integro-differential equation \eqref{smct} becomes
\begin{equation}
\tilde{c}(s)=[s+i\Delta_{0}+\int_{\zeta_\text{min}}^{\zeta_\text{max}} d\zeta\frac{J(\zeta)}{s+i\zeta}]^{-1}\label{sminvs}
\end{equation}under a Laplace transform.
Then $c(t)$ is obtainable by applying an inverse Laplace transform
\begin{equation}
c(t)=\frac{1}{2\pi i}\int_{\sigma-i\infty}^{\sigma+i\infty}\tilde{c}(s)e^{st}ds\label{smlap}
\end{equation} to Eq. \eqref{sminvs}.
Equation \eqref{smlap} is evaluated by using the the residue theorem. The residue is contributed by the poles of $\tilde{c}(s)$, which is found via
\begin{equation}
   E/\hbar=\Delta_{0}+\int_{\zeta_\text{min}}^{\zeta_\text{max}} d\zeta \frac{J(\zeta)}{E/\hbar-\zeta}\equiv Y(E),\label{smpeq}
\end{equation}where $s=-iE/\hbar$. Equation \eqref{smpeq} has at most one isolated pole in the regime either $E\in(-\infty,\zeta_\text{min}]$ or $E\in[\zeta_\text{max},+\infty)$ provided $Y(\hbar\zeta_\text{min})<\hbar\zeta_\text{min}$ or $Y(\hbar\zeta_\text{max})>\hbar\zeta_\text{max}$, respectively. Using the residue theorem, we have
\begin{equation}
    c(t)=\sum_{j=1}^MZ_je^{{-i\over\hbar}E^b_jt}+\int_{\zeta_\text{min}}^{\zeta_\text{max}}\Theta(E)e^{-iEt}dE,\label{smctr}
\end{equation}where $\Theta(E)={J(E)\over[E-\Delta_0-\Upsilon(E/\hbar)]^2+[\pi J(E)]^2}$, $M$ being the number of the bound states, and $Z_{j}=[1+\int_{\zeta_\text{min}}^{\zeta_\text{max}}\frac{J(\zeta)d\zeta}{(E^\text{b}_{j}/\hbar-\zeta)^2}]^{-1}$ is the residue contributed by the $j$th bound state. Oscillating with time in continuously changing frequencies ${E/\hbar}$ of the energy band, the integrand tends to zero in the long-time limit due to the out-of-phase interference. Thus, the steady-state solution of Eq. \eqref{smctr} is
\begin{equation}
\lim_{t\rightarrow\infty}c(t)=\begin{cases}0, ~~~~~~~~~~~~~~~~~~~~\text{no bound state}\\ \sum_{j=1}^MZ_je^{{-i\over\hbar}E^\text{b}_jt}, ~M~\text{bound states}\end{cases}\hspace{-.3cm}.
\end{equation}

\bibliography{Ref}